\documentclass[aps,pra,twocolumn,superscriptaddress,groupedaddress,longbibliography,showpacs]{revtex4-1}  

\usepackage{graphicx}  
\usepackage{bm}        
\usepackage{amssymb}   
\usepackage{amsmath}
\usepackage{amsfonts}
\usepackage{natbib}
\usepackage{float}
\usepackage{xcolor}
\usepackage{ulem}
\usepackage{txfonts}
\usepackage{times}
\usepackage{subfigure}
\usepackage{mathrsfs}
\usepackage{hyperref}
\usepackage{bbold}
\usepackage{comment}
\usepackage{fancyhdr}
\usepackage{ulem}
\usepackage{enumerate}

\let\oldhat\hat
\renewcommand{\hat}[1]{\oldhat{\mathbf{#1}}}

\newcommand{\beq}{\begin{equation}}
\newcommand{\eeq}{\end{equation}}

\newcommand{\vQ}{\bm{\mathit Q}}
\newcommand{\vq}{\mathbf{q}}
\newcommand{\qq}{\mathbf{q}}

\newcommand{\iu}{{i\mkern1mu}}

\newcommand{\eps}{\varepsilon}

\newcommand{\gs}{\text{g.s.}}

\newcommand{\FeTeSe}{Fe(Te$_{1-x}$Se$_{x}$)}

\begin{document}

\title{Unified Spin Model for Magnetic Excitations in Iron Chalcogenides}
\author{Patricia Bilbao Ergueta}
\author{Wen-Jun Hu}
\author{Andriy H. Nevidomskyy}
\affiliation{Department of Physics and Astronomy, Rice University, Houston, Texas 77005, USA}
\pacs{
75.10.-b	 
74.70.Xa, 
74.25.-q 
}

\begin{abstract}
Recent inelastic neutron scattering (INS) measurements on FeSe and Fe(Te$_{1-x}$Se$_x$) have sparked intense debate over the nature of the ground state in these materials. Here we propose an effective bilinear-biquadratic spin model, which is shown to consistently describe the evolution of low-energy spin excitations in FeSe, both under applied pressure and upon Se/Te substitution.
 The phase diagram, studied using a combination of variational mean-field, flavor-wave calculations and density-matrix renormalization group (DMRG), exhibits a sequence of transitions between the columnar antiferromagnet common to the iron pnictides, the  nonmagnetic ferroquadrupolar phase attributed to FeSe, and the double-stripe antiferromagnetic order known to exist in Fe$_{1+y}$Te. The calculated spin structure factor in these phases mimics closely that observed with INS in the Fe(Te$_{1-x}$Se$_x$), series. In addition to the experimentally established phases, the possibility of incommensurate magnetic order is also predicted.
\end{abstract}

\maketitle

\section{Introduction}
Iron chalcogenides are considered to be the most correlated of the iron-based family of superconductors and have been the subject of intensive study, both theoretically and experimentally. In Fe$_{1+y}$Te, it was found early on that the magnetic ground state has  an unusual double-stripe (DS) structure characterized by the ordering wave vector $\bm{Q}=(\pi/2,\pi/2)$ in the one-iron unit-cell notation~\cite{Bao2009,Li2009,Wen2009}. This state is in stark contrast to the parent compounds of iron pnictides that have a columnar antiferromagnetic (CAFM) ground state~\cite{LaCruz2008,Lumsden2010,Dai2015}. Upon doping with selenium, the DS magnetism disappears, resulting in a nonmagnetic ground state in \FeTeSe\, (for sufficiently large $x$) \cite{Qiu2009, Lee2010, Lumsden_NPhys2010,Liu2010,Xu2016}. The nature of this state, extending all the way to the stoichiometric FeSe, has been the subject of intense debate recently, with elastic neutron scattering showing no sign of magnetic Bragg peaks in FeSe~\cite{McQueen2009,Bendele2010}. The INS studies have found large finite-energy spectral weight at wave vectors $\bm{Q}_{1,2} = (\pi,0)/(0,\pi)$ \cite{Rahn2015, QWang2015, QWang2016,Shamoto2015}, which are characteristic of the CAFM magnetic order in the iron pnictides~\cite{LaCruz2008}. This suggests that FeSe is close to magnetic ordering, presumably to the CAFM phase. Indeed, it was shown that magnetism can be reached by applying hydrostatic pressure to FeSe, as indicated by the recent transport, ac susceptibility, x-ray scattering, and NMR measurements~\cite{Bendele2012,Terashima2015,Kothapalli2016,Wang-NMR2016}. 

The conspicuous lack of magnetic ordering under ambient pressure in FeSe has led to several theoretical proposals for the unusual nature of the ground state in this material~\cite{Glasbrenner2015,Wang2015,Yu2015,ZWang2016}. For the theory to be consistent, it is important that it should be able to describe not only the lack of magnetic ordering in FeSe, but also the appearance of magnetism under applied pressure and with Te doping. In this paper, we show that the recently proposed theory of the spin ferroquadrupolar (FQ) ground state for FeSe~\cite{ZWang2016} indeed satisfies these requirements  and successfully describes the evolution of the INS data as a function of Te doping in \FeTeSe, in qualitative accord with the recent INS experiments~\cite{Xu2016}.
 
In this work, we use the frustrated bilinear-biquadratic \mbox{spin-1} Heisenberg model as a basis, which has been employed by many authors to model iron pnictides and chalcogenides~\cite{Fang2008,Wysocki2011,Yu2012,Wang2015,Yu2015,ZWang2016,Lai2016,Hu2016}:
\begin{equation}\label{eq:model}
\mathcal{H}= \frac{1}{2} \sum_{i,j} J_{ij} \bm{S}_i \cdot \bm{S}_j +  \frac{1}{2}  \sum_{i,j} K_{ij} (\bm{S}_i \cdot \bm{S}_j )^2,
\end{equation}
where $\bm{S}_i$ is the quantum spin-1 operator on site $i$, describing the Hund's-coupled spins of conduction electrons in the half-filled Fe $d_{xz}$ and $d_{yz}$ orbitals. Earlier studies \cite{Wysocki2011,Yu2012,ZWang2016} have proposed that it is sufficient to limit the spatial extent of the interactions to the first and second nearest neighbors, $J_{ij}=\{J_1, J_2\},\,K_{ij}=\{ K_1, K_2\}$, in order to model the INS data on the iron pnictides and FeSe.  
 In this paper, we show that including the third-neighbor Heisenberg interaction $J_3$ is necessary to describe the DS magnetic state of Fe$_{1+y}$Te and that the increasing $J_3$ strength describes qualitatively the effect of Te doping in \FeTeSe. Using the variational mean-field, flavor-wave expansion and the DMRG calculations, we compute the phase diagram and establish that the evolution of the calculated dynamic spin-structure factor $S(\vq,\omega)$ with increasing $J_3$ mimics that observed in INS data in \FeTeSe\cite{Xu2016}. Crucially, the obtained phase diagram 
 naturally describes both this evolution and the tendency towards the CAFM ordering under the applied pressure in FeSe \cite{Bendele2012,Terashima2015,Kothapalli2016,Wang-NMR2016} within a single unified theory. This $J_1$-$J_2$-$J_3$-$K_1$-$K_2$ theory, shown earlier to describe semiquantitatively the spin dynamics of BaFe$_2$As$_2$ iron pnictides with very few fitting parameters~\cite{Wysocki2011,Yu2012}, can thus be considered an effective spin model of both iron pnictides and chalcogenides, and is therefore of fundamental importance to the field of iron-based superconductors. Of course one can attempt to include third-neighbor biquadratic ($K_3$) and farther interactions; however, the predictions of the present model readily agree with the INS results and guided by Occam's razor, we therefore propose that the interactions up to \{$J_3$, $K_2$\} order be considered sufficient.

This paper is organized as follows. The analytical approaches, namely, the variational mean-field and flavor-wave techniques, are introduced in Sec.~\ref{sec:approaches}, with the respective calculated phase diagrams presented in Sec.~\ref{sec:phd}. Our conclusions are corroborated with the state-of-the-art DMRG calculations, summarized in Sec.~\ref{sec:DMRG}. We proceed to calculate the dynamical spin-structure factors and provide detailed comparison with the INS experiments on \FeTeSe\, in Sec.~\ref{sec:Sqw}, before exploring the theoretical indications of the incommensurate magnetic order in Sec.~\ref{sec:IC}. Finally, we conclude with the discussion and outlook in Sec.~\ref{sec:conclusions}.

\section{Analytical approaches}\label{sec:approaches}
\subsection{Variational Mean field}\label{sec:MF}

\begin{figure*}[!tb]
\includegraphics[width=1\textwidth]{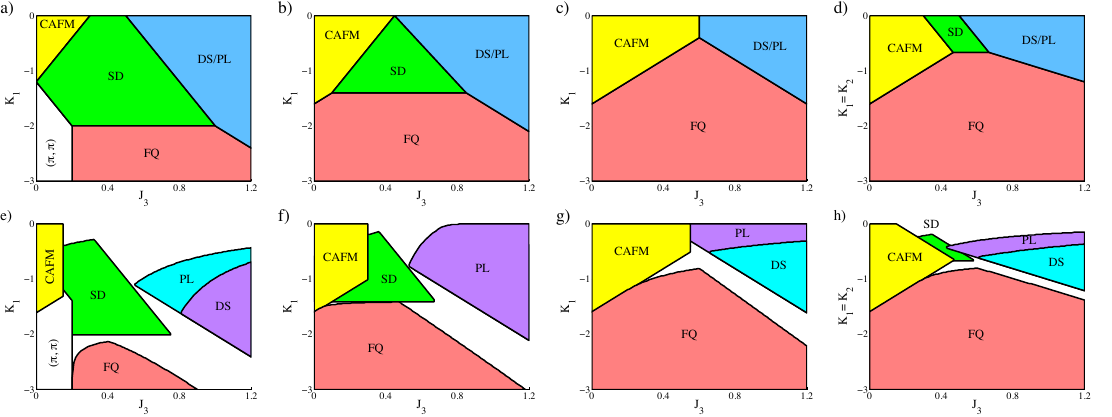}
\caption{Phase diagram of the Hamiltonian Eq.~\eqref{eq:model} with $J_1=1,J_2=0.8$ and periodic boundary condition on a $4 \times 4$  cluster as a function of $J_3$ and $K_1$ for (a, e) $K_2=0$, (b, f) $K_2=-0.3$, (c, g) $K_2=-0.8$, and (d, h) $K_2=K_1\equiv K$. Panels (a)-(d) were obtained within variational mean-field calculation. The effect of spin-dipolar and quadrupolar fluctuations on the phase diagram is shown in panels (e)-(h) by flavor-wave calculation. The white regions show the regime of parameters where the flavor-wave expansion is unstable, indicating likely incommensurate spin order.}
\label{Fig1}
\end{figure*}

In FeTe, attempts to fit the experimental spin-wave dispersion with a purely Heisenberg model required highly anisotropic exchange couplings $J_{1a}\neq J_{1b}$~\cite{Lipscombe2011}. In fact, both of them were required to be ferromagnetic~\cite{Lipscombe2011}, in contrast to all the iron pnictides where the antiferromagnetic superexchange is necessary~\cite{Dai2015}.  Below we show that including the biquadratic spin-spin interaction $K_{ij}$ makes it possible to obtain the experimentally observed DS phase (also referred to as bicollinear phase in the literature) with the ordering wave vector $\bm{Q}=(\pi/2,\pi/2)$ while maintaining an isotropic nearest-neighbor (NN) exchange, as shown in the phase diagram in Fig.~\ref{Fig1}. We note that a similar problem arises when attempting to fit the high-energy spin-wave dispersion in the parent compounds of the iron pnictides in the CAFM phase~\cite{Zhao2009, Harriger2011}, 
and it was proposed by us and others that this problem can be resolved by inclusion of a suitable NN biquadratic interaction $K_1<0$~\cite{Wysocki2011,Yu2012}. 

It is useful to recast the $J_{ij}-K_{ij}$ model in terms of 
the traceless symmetric quadrupolar tensor:
\beq 
Q^{\alpha\beta} = S^\alpha S^\beta +S^\beta S^\alpha-\frac{2}{3} S(S+1) \delta_{\alpha \beta},\label{eq:Q}
\eeq
whose five independent components are convenient to cast into a five-component vector \mbox{$\bm{\mathit Q} \equiv \left[(Q^{xx}-Q^{yy})/2, (2Q^{zz}-Q^{xx}-Q^{yy})/2\sqrt{3}, Q^{xy}, Q^{yz}, Q^{xz}\right]$}. Using an identity $2 (\bm{S}_i \cdot \bm{S}_j)^2 = \vQ_i \cdot \vQ_j -\bm{S}_i \cdot \bm{S}_j+\frac{8}{3}$ for $S=1$, the model in Eq.~(\ref{eq:model}) can then be rewritten in the form
\begin{equation}\label{eq:model2}
\mathcal{H}=\frac{1}{2} \sum_{i,j} \left(J_{ij}-\frac{K_{ij}}{2}\right) \bm{S}_i \cdot \bm{S}_j + \frac{1}{4} \sum_{i,j} K_{ij} \left(\bm{\mathit Q}_i \cdot \bm{\mathit Q}_j + \frac{8}{3}\right).
\end{equation}

In order to get an insight into the properties of this model, we first seek a mean-field solution, which is equivalent to writing the wave function in a separable form,
\beq
|\Psi\rangle_\text{MF} = \prod_i |\vec{d}_i\rangle , \label{eq:MF}
\eeq
 in terms of the single-particle states $| \vec{d}_i\rangle=\sum_\alpha d_i^\alpha |\alpha \rangle $~\cite{Papanicolaou1988, ZWang2016}. Anticipating the magnetic as well as quadrupolar solutions, it is convenient to use a quadrupolar basis of time-reversal invariant states $|\alpha\rangle = \{ \, |x\rangle,\, |y\rangle,\,  |z\rangle \, \}$, which are linear superpositions of the familiar $|S_z\rangle$ states:
 
 \begin{equation}
 |x\rangle=\iu \frac{|1\rangle-|\bar{1}\rangle}{\sqrt{2}}, \qquad
 |y\rangle=\frac{|1\rangle+|\bar{1}\rangle}{\sqrt{2}}, \qquad
 |z\rangle=-\iu |0\rangle.
 \end{equation}
The spin operators transform accordingly and can be written conveniently in the following form:
 \beq
 S^\nu = -i\eps_{\nu\gamma\delta} |\gamma\rangle  \langle \delta|. \label{eq:spin}
 \eeq
 
\noindent 
Similarly, the quadrupolar operators $Q^{\alpha\beta}$ introduced in Eq.~(\ref{eq:Q}) take on the following form in this basis:
\beq
Q_{\alpha\beta} = \frac{2}{3}\delta_{\alpha\beta} - |\beta\rangle\langle\alpha| - |\alpha\rangle\langle\beta|. \label{eq:Q2}
\eeq
Using Eqs.~(\ref{eq:spin}) and (\ref{eq:Q2}), we can now evaluate the energy of the Hamiltonian in Eq.~(\ref{eq:model2}) in the mean-field ansatz given by the choice of directors $|\vec{d}_i\rangle=\sum_\alpha d_i^\alpha|\alpha\rangle$ in Eq.~(\ref{eq:MF}). This results in the following mean-field expression for the energy:
\begin{equation}\label{eq:E0}
E_0 = \frac{1}{2N} \sum_{i,j} \left[ J_{ij} |\langle  \vec{d}_i | \vec{d}_j \rangle |^2 - (J_{ij}-K_{ij}) |\langle  \vec{d}_i | \vec{d}_j^* \rangle |^2 +K_{ij} \right].
\end{equation}

\noindent
We then perform a variational search by minimizing Eq.~\eqref{eq:E0} with respect to the set of directors $\{\vec{d}_i\}$ restricted to a supercell of lattice vectors. Choosing a larger supercell allows one to consider the states that do not preserve translational symmetry of the lattice, such as  staggered spin or quadrupolar orders. For the purpose of this work, it was sufficient to choose a supercell of dimension $4 \times 4$ with periodic boundary conditions.

We note that the directors $\vec{d}_i=\vec{u}_i+i\vec{v}_i$ are complex three-component objects satisfying the constraints $|\vec{u}_i|^2 + |\vec{v}_i|^2=1$ and $\vec{u}_i\cdot\vec{v}_i=0$. It follows from Eq.~(\ref{eq:spin}) that the expectation value of spin is
\beq
\langle \vec{d}|\bm{S}|\vec{d}\rangle = 2 \vec{u}\times \vec{v},
\eeq
so that the long-range dipolar order is only present when both $\vec{u}$ and $\vec{v}$ are nonzero, whereas the pure quadrupolar states are identified by $\langle\bm{S}\rangle=0$ and correspond to the director $\vec{d}$ being purely real or purely imaginary.

\begin{figure}[!tb]
\includegraphics[width=0.48\textwidth]{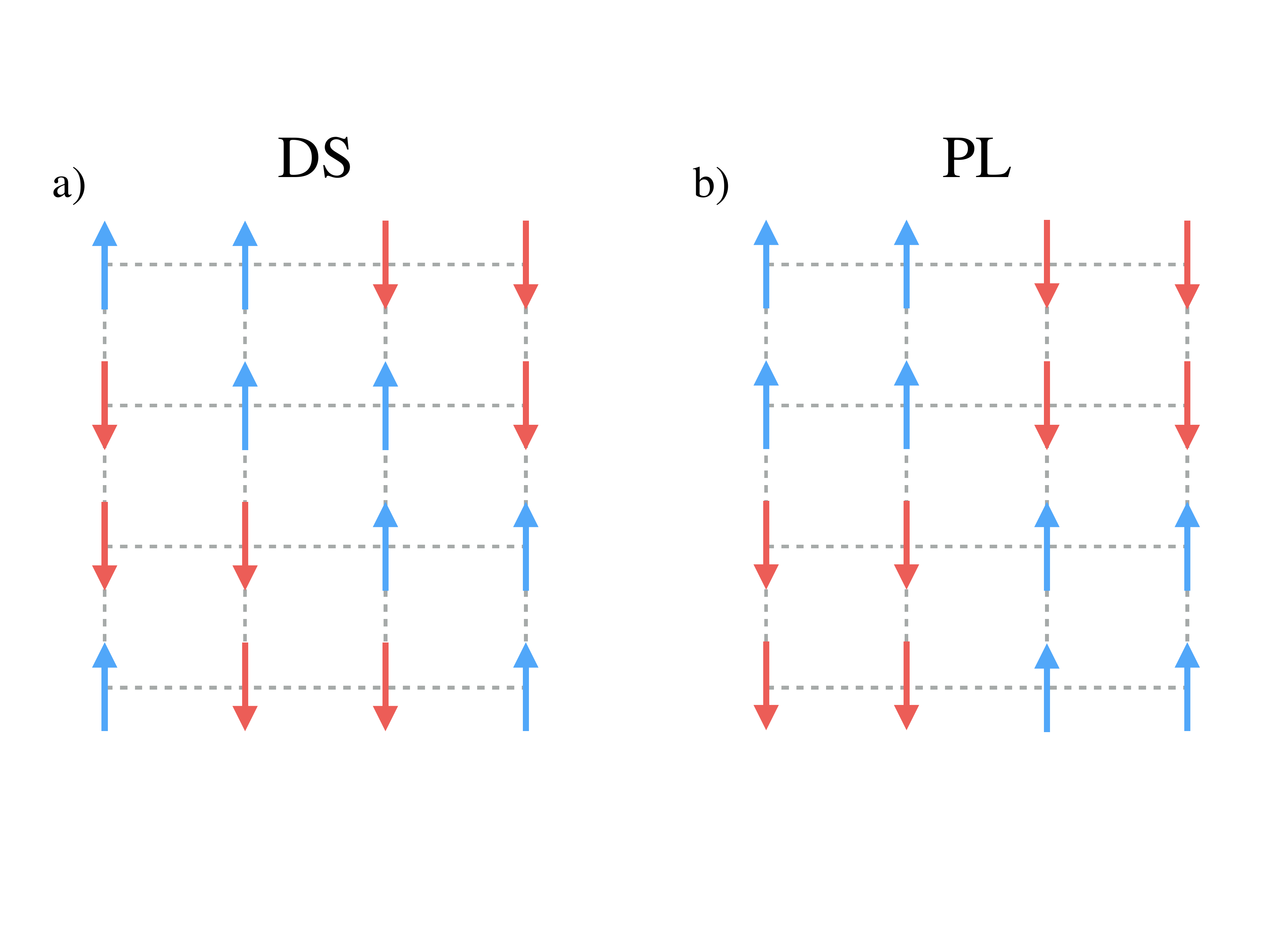}
\caption{Schematic depiction of (a) bicollinear DS state and (b) PL state.}\label{Fig:DS}
\end{figure}

Depending on the parameter regime, we find five magnetically ordered phases:
\begin{enumerate}[(i)]
 \item CAFM, with ordering wave vector $\bm{Q}=(\pi,0)/(0,\pi)$; 
 \item N\'eel state with $\bm{Q}=(\pi,\pi)$;
 \item DS state with $\bm{Q}=\pm(\pi/2,\pi/2)$, depicted in Fig.~\ref{Fig:DS}(a);
 \item Plaquette (PL) state, with $\bm{Q}=(\pm \pi/2, \pm \pi/2)$, depicted in Fig.~\ref{Fig:DS}(b);
 \item Staggered dimer (SD) state depicted in Fig.~\ref{Fig:SD}(a), identified by $\bm{Q}=(\pm \pi/2,\pi)/(\pi,\pm \pi/2)$.
\end{enumerate}

In addition, we also find an extensive region of the nonmagnetic FQ phase, characterized by a uniform set of directors $\vec{d}_i=\vec{d} \; \forall i$, with a vanishing magnetic (dipolar) moment: $0=\langle \bm{S}_i \rangle \equiv 2\, \text{Re}[ \vec{d}_i ] \times \text{Im}[ \vec{d}_i ]$. 
This is the only stable nonmagnetic state in the phase diagram for the studied parameter regime and is natural to interpret as the ground state of FeSe, as was shown in Ref.~\onlinecite{ZWang2016}. This interpretation is further strengthened by a very good accord between the theoretical spin-structure factors~\cite{ZWang2016} and those measured in INS experiments~\cite{QWang2015,Shamoto2015}. 

The mean-field energies of the aforementioned phases are as follows:
\begin{subequations}
	\begin{align}
		E_\text{CAFM}   &= -2J_2 + 2J_3 +3K_1 +4K_2, \\
		E_\text{N\'eel} &= -2J_1 + 2J_2 +2J_3 +4K_1 + 2K_2, \\
		E_\text{DS/PL}     &= 3K_1 +3K_2 -2J_3, \label{eq:E_DS} \\ 
		E_\text{SD} &= -J_1 + \frac{7}{2}K_1 + 3K_2, \\
		E_\text{FQ}     &= 4K_1 + 4K_2.
	\end{align}
\end{subequations}
The resulting mean-field phase diagrams, shown in Fig.~\ref{Fig1}, will be discussed later in Sec.~\ref{sec:phd}. 
We note that within the variational mean-field method, the bicollinear DS phase [Fig.~\ref{Fig:DS}(a)] is degenerate in energy with the PL state depicted  in Fig.~\ref{Fig:DS}(b). We shall comment further on the distinction between these two states when discussing the phase diagram results in Sec.~\ref{sec:phd} and the DMRG results in Sec.~\ref{sec:DMRG}.

\subsection{Fluctuations around mean-field: Flavor wave expansion}\label{sec:flwave}
In order to improve on the mean-field solution, we have performed a series of flavor-wave calculations, which accounts for the fluctuations in the spin-dipolar as well as spin-quadrupolar channels~\cite{Papanicolaou1988, Tsunetsugu2006, Lauchli2006, Muniz2014}.
The essence of this technique consists in expanding the local operators $\mathscr{O}_i$ in terms of the three species $(\alpha,\beta=\{0,1,2\})$ of bosons that transform in the fundamental representation of group SU(3):

\begin{equation}
	\mathscr{O}_i = \sum_{\alpha \beta} b_{i,\alpha}^\dagger O_i^{\alpha \beta} b_{i,\beta}, \quad 
	\sum_\alpha b_{i,\alpha}^\dagger b_{i,\alpha}=1.
\end{equation}



In a phase with long-range order (including quadrupolar orders), some linear combination of bosons is condensed, $\sum_\alpha \langle \mathcal{V}_i^{\alpha 0}b_{i,\alpha}^\dagger \rangle \equiv \langle \tilde{b}_{i,0}^\dagger \rangle  \neq 0$. This can be cast in terms of a unitary transformation into a new basis:

\begin{subequations}
	\begin{align}
		\tilde{\bm{b}}_{i} &= \mathcal{V}_i^\dagger \bm{b}_i, \label{Eq.V-matrix}\\
		\tilde{O}_i &= \mathcal{V}_i^\dagger O_i \mathcal{V}_i,
	\end{align}
\end{subequations}

 where the matrix form of $\mathcal{V}_i$ is determined by the mean-field ground state in Eq.~(\ref{eq:MF}), expressed by an appropriate choice of directors $|\vec{d}_i\rangle$.

Below, we outline the key steps in the flavor-wave procedure, while relegating further details to the Appendix:
\begin{enumerate}[1)]
\item For a given mean-field ansatz $|\vec{d}_i\rangle$, determine the unitary matrices $\mathcal{V}_i$ (for each sublattice $i$);
\item Condense the appropriate boson with the local constraint by writing $\tilde{b}_{i,0}=\sqrt{1-\tilde{b}_{i,1}^\dagger \tilde{b}_{i,1}-\tilde{b}_{i,2}^\dagger \tilde{b}_{i,2}}$;
\item Expand the square roots in the Hamiltonian Eq.~(\ref{eq:model2}) up to quadratic order in $\tilde{b}_{i,a}^\dagger$ and $\tilde{b}_{i,a}$ ($a=1,2$);
\item Diagonalize the resulting expression, using the Bogoliubov transformation, to obtain the flavor-wave Hamiltonian in terms of new bosonic operators $\{\alpha_{\bm{q},\nu}^\dagger,\alpha_{\bm{q},\nu} \}$:
\beq
\mathcal{H}_{\text{fw}}=\sum_{\bm{q}}\sum_{\nu} \omega_{\bm{q},\nu} (\alpha_{\bm{q},\nu}^\dagger \alpha_{\bm{q},\nu}+1/2) + N \cdot \text{const},
\eeq 
where $\nu$ denotes different excitation branches.
\end{enumerate}

The contribution of the zero-point fluctuations, 
\beq
E_{zp} = \frac{1}{2N}\sum_{\bm{q},\nu} \omega_{\bm{q},\nu} + \text{const},
\eeq
 lowers the energy compared to the mean-field value, resulting in the shift of the phase boundaries, as seen in Figs.~\ref{Fig1}(e)-\ref{Fig1}(h). As we shall see in the following section, in certain cases (especially near the phase boundaries) the mean-field solution turns out to be unstable, as evidenced by the softening in the spectrum of flavor-wave excitations, which then acquire an imaginary component. 
 At this point, the mean-field solution is not to be trusted, and a different method (such as DMRG) must be used to establish the nature of the ground state, as we discuss in Sec.~\ref{sec:DMRG}. Nevertheless, we shall demonstrate in Sec.~\ref{sec:IC} that even when the mean-field solution turns out to be unstable, the analysis of the flavor-wave spectrum softening allows one to glean further information into the nature of the resulting ground state, including the possibility of incommensurate order. 
 


\section{Phase diagrams}\label{sec:phd}

A representative mean-field phase diagram is shown in Figs.~\ref{Fig1}(a)-\ref{Fig1}(d) for four different cases: $K_2=0$, $K_2=-0.3$, $K_2=-0.8$, and $K_2=K_1\equiv K$, respectively. We have chosen the units such that $J_1=1$ and further fixed $J_2=0.8$ in accord with the \textit{ab initio} calculations~\cite{Wysocki2011}. This leaves $J_3$ and $K_1, K_2$ as free parameters in the calculations. In this article, we focus on negative values of $K_1$ and $K_2$, as those are obtained by fitting the INS spectra to the $J_{ij}-K_{ij}$ model~\cite{Wysocki2011,Yu2012} and as it has also been shown that positive values lead to unwanted phases~\cite{ZWang2016}. We also note that the large negative $K_1$ is also expected from the spin crossover model by Chaloupka and Khaliullin~\cite{Chaloupka2013}, and large $|K_1|$ also naturally arises within the Kugel-Khomskii type models when the orbitals order inside the nematic phase~\cite{Ergueta2015}.
 
\begin{figure}[!tb]
\includegraphics[width=0.48\textwidth]{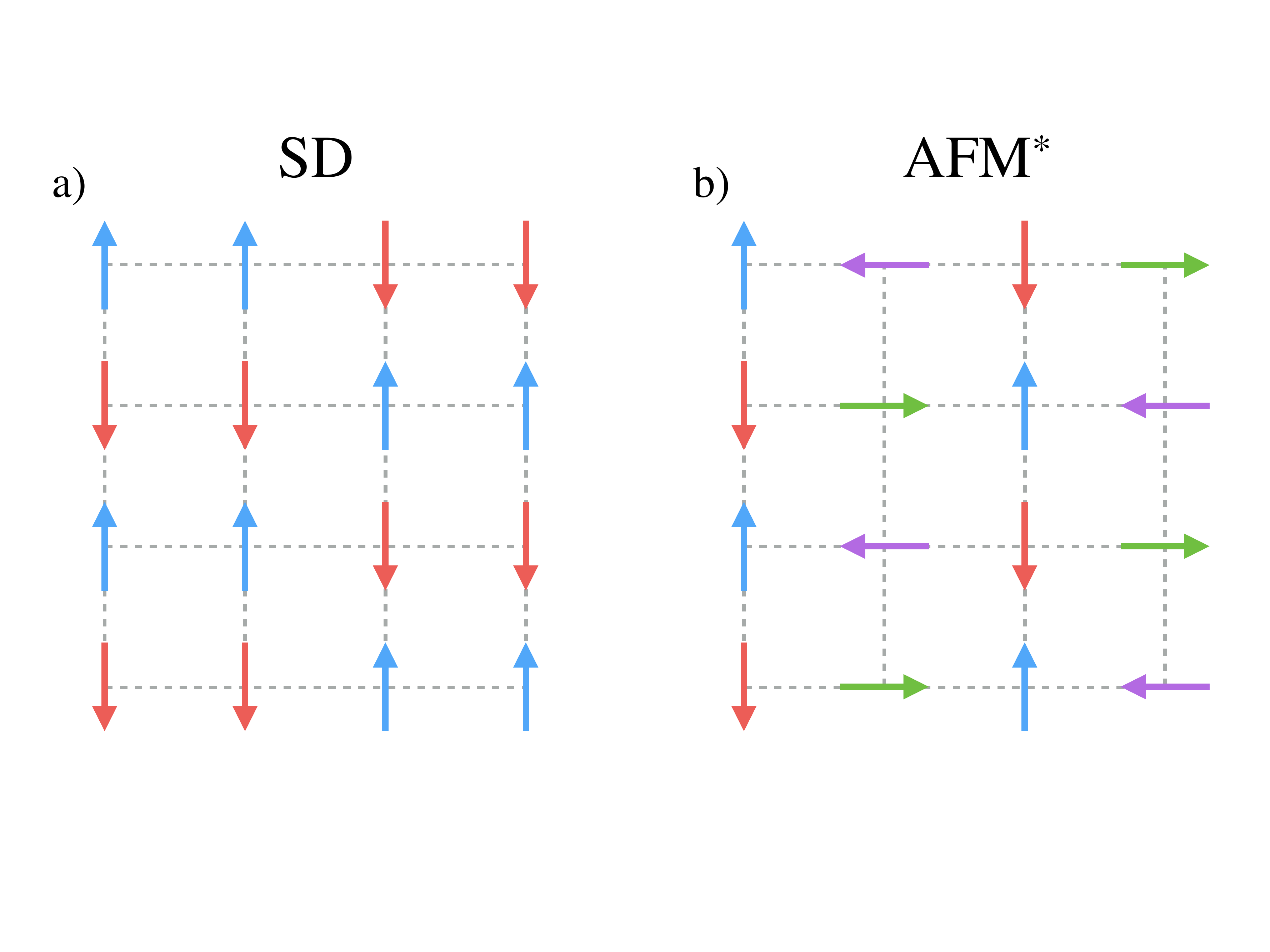}
\caption{Schematic depiction of (a) staggered-dimer (SD) state and (b) AFM$^\ast$ $(\pi/2,\pi)$ state introduced in Ref.~\cite{Hu2016}. We find the two states to always be degenerate in the entire studied parameter regime presented in this paper.\label{Fig:SD}}
\end{figure}

 As Fig.~\ref{Fig1} indicates, the CAFM phase dominates for small $J_3$, provided $|K_1|$ is not too large, while for sufficiently negative $K_1$ we observe the appearance of either the FQ or the $(\pi,\pi)$ N\'eel phase. This is due to the fact that in the absence of $K_2$, a negative biquadratic coupling $K_1$ renormalizes the NN Heisenberg interaction, making the $(\pi,\pi)$ correlations stronger~\cite{Yu2012,Ergueta2015}. Since the N\'eel phase has not been observed in either iron pnictides or chalcogenides, our calculations support the conclusion that $K_2$ must be present and negative. Above a certain critical value of $K_1$, the FQ order is stabilized and a direct transition between the FQ and CAFM phases is achieved~\cite{ZWang2016}, mimicking the experimentally observed transition from the nonmagnetic to the antiferromagnetic state in FeSe under applied pressure~\cite{Bendele2012,Terashima2015,Kothapalli2016,Wang-NMR2016}.
For sufficiently large $J_3$, a DS magnetic order is obtained in Fig.~\ref{Fig1}, which is well established in Fe$_{1+y}$Te~\cite{Bao2009,Li2009,Wen2009}. An intermediate SD phase (colored green in Fig.~\ref{Fig1}) also typically appears between the CAFM and DS or PL phases [although there is a parameter regime where it is absent, see Figs.~\ref{Fig1}(c) and \ref{Fig1}(g)]. This phase, 
characterized by wavevectors $(\pi,\pm\pi/2)$ or $(\pm\pi/2,\pi)$, 
breaks the lattice $C_4$ symmetry and is depicted schematically in Fig.~\ref{Fig:SD}(a).  There may be indirect experimental evidence of such an intermediate phase in FeSe under applied pressure~\cite{Wang-NMR2016}. We note that the SD phase is distinct from the so-called AFM$^\ast$ $(\pi,\pi/2)$ phase studied in Ref.~\cite{Hu2016} [see Fig.~\ref{Fig:SD}(b)]; within the mean-field treatment, we find both phases to be degenerate in the entire parameter regime presented in this paper.

%
%

As depicted in Figs.~\ref{Fig1}(a)-(d), the DS and PL phases are exactly degenerate at the mean-field level. However, quantum fluctuations, captured within the flavor-wave expansion, lift the degeneracy so that one or the other phase becomes the true ground state, depending on the region of the parameter regime. 
For small values of $K_2$ [see Fig.~\ref{Fig1}(e)], we find that the PL phase is the ground state within its region of stability. Outside of this region, fluctuations destroy the PL order and the DS phase is stabilized instead over a wider parameter range. For larger values of $K_2$ [see Fig.~\ref{Fig1}(g)], the behavior is the opposite, with the DS phase being lower in energy when both phases are possible but the PL phase remaining stable in the wider parameter regime. 
Figure~\ref{Fig1}(f) shows the PL phase to always be the ground state for $K_2=-0.3$. However, the energy differences between the PL and DS phases are in this case the smallest out of all the cases we studied and their stability regions almost overlap. 
The $K_2=K_1$ case [see Fig.~\ref{Fig1}(h)] showcases the same behavior that is observed for the larger values of $K_2$ when it comes to the PL/DS regions. Additionally, we find that taking into account the effect of quantum fluctuations greatly reduces the region of stability of the SD phase (colored green) compared to the mean-field results in Fig.~\ref{Fig1}(d).

\begin{figure}[!t]
\includegraphics[width=0.50\textwidth]{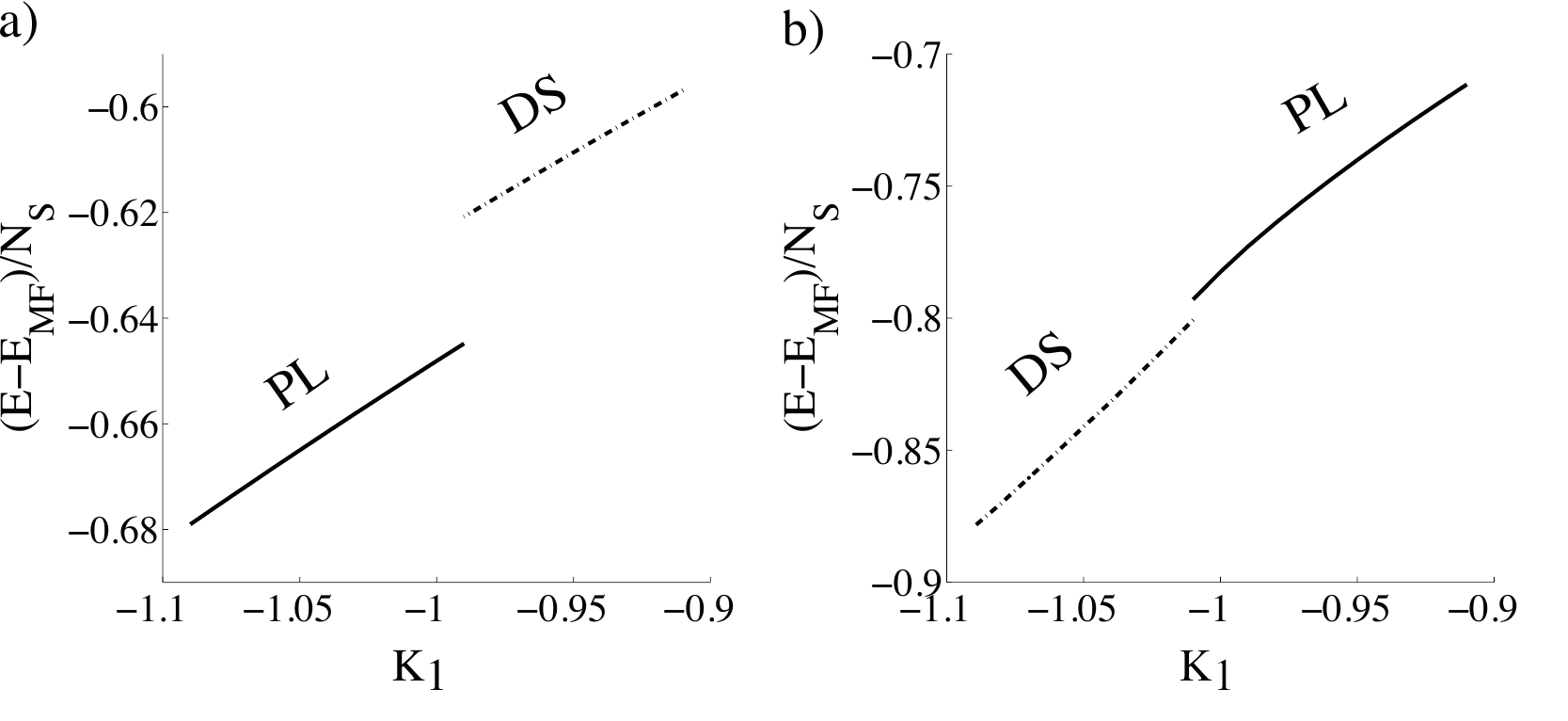}
\caption{The zero-point energies of the PL (solid line) and DS (dashed line) phases, including the contribution from flavor-wave fluctuations, plotted across a constant $J_3=1.0$ cut through the phase boundary between the two phases, for two different values of $K_2$: (a) $K_2=0$ and (b) $K_2=K_1$. A first-order jump in energy is observed at the transition, more pronounced for small $|K_2|$ as in panel (a).}
\label{energies}
\end{figure}

Due to the stability regions being different for the PL and DS phases, there is a first-order discontinuity in the energy at the phase boundary between the two.
 This is shown in Fig.~\ref{energies} for the two cases where this jump is most appreciable. For the rest of the cases, the energy difference is even smaller and vanishes when the phase boundaries approach one another. 
 The $K_2=-0.3$ case [see Fig.~\ref{Fig1}(f)] is a good example thereof, with the PL phase being only slightly lower in energy than the DS phase, and the two phase boundaries almost overlapping.

As mentioned earlier in Sec.~\ref{sec:flwave}, the flavor waves may result in unstable regions near the mean-field phase boundary between different phases. Physically, this means that order-parameter fluctuations destroy the given long-range order, indicating the tendency of the systems towards a different ground state. Such unstable solutions are indicated by the white unshaded regions in Figs.~\ref{Fig1}(e)-(h). 
Besides the appearance of these unstable regions, the qualitative behavior of the phases remains the same, with only the numerical values of the phase boundaries shifting with respect to their mean-field values. The flavor-wave expansion is nevertheless very valuable for two reasons: first, it allows for the calculation of the dynamical quantities, such as spin-structure factor which will be discussed in Sec.~\ref{sec:Sqw}, and second, the details of the instability in the flavor-wave spectrum provide clues as to the origin of the true ground state, as we shall explore in Sec.~\ref{sec:IC}.

\section{DMRG solution}\label{sec:DMRG}

Having established the mean-field phase diagram, we verify the stability of the phases shown in Fig.~\ref{Fig1} using unbiased SU(2) DMRG calculations~\cite{White1992,mcculloch2002,gong2014square,gong2014kagome} on $L \times 2L$ rectangular cylinders with $L=(4,6,8)$~\footnote{$L$ represents the size of y-direction which has periodic boundary condition}.
 We keep up to $4000$ SU$(2)$ states, leading to truncation errors around $10^{-5}$.
The finite-size analysis for the CAFM and FQ phases is identical to that performed in Ref.~\cite{ZWang2016} so we only show the results on the largest cylinder ($L=8$ unless noted otherwise), taking a horizontal cut at $K_1=K_2\equiv K=-0.3$ in the phase diagrams in Figs.~\ref{Fig1}(d) and \ref{Fig1}(h) and studying the effect of increasing $J_3$.

\begin{figure}
	\begin{center}
		\includegraphics[width=0.75\columnwidth]{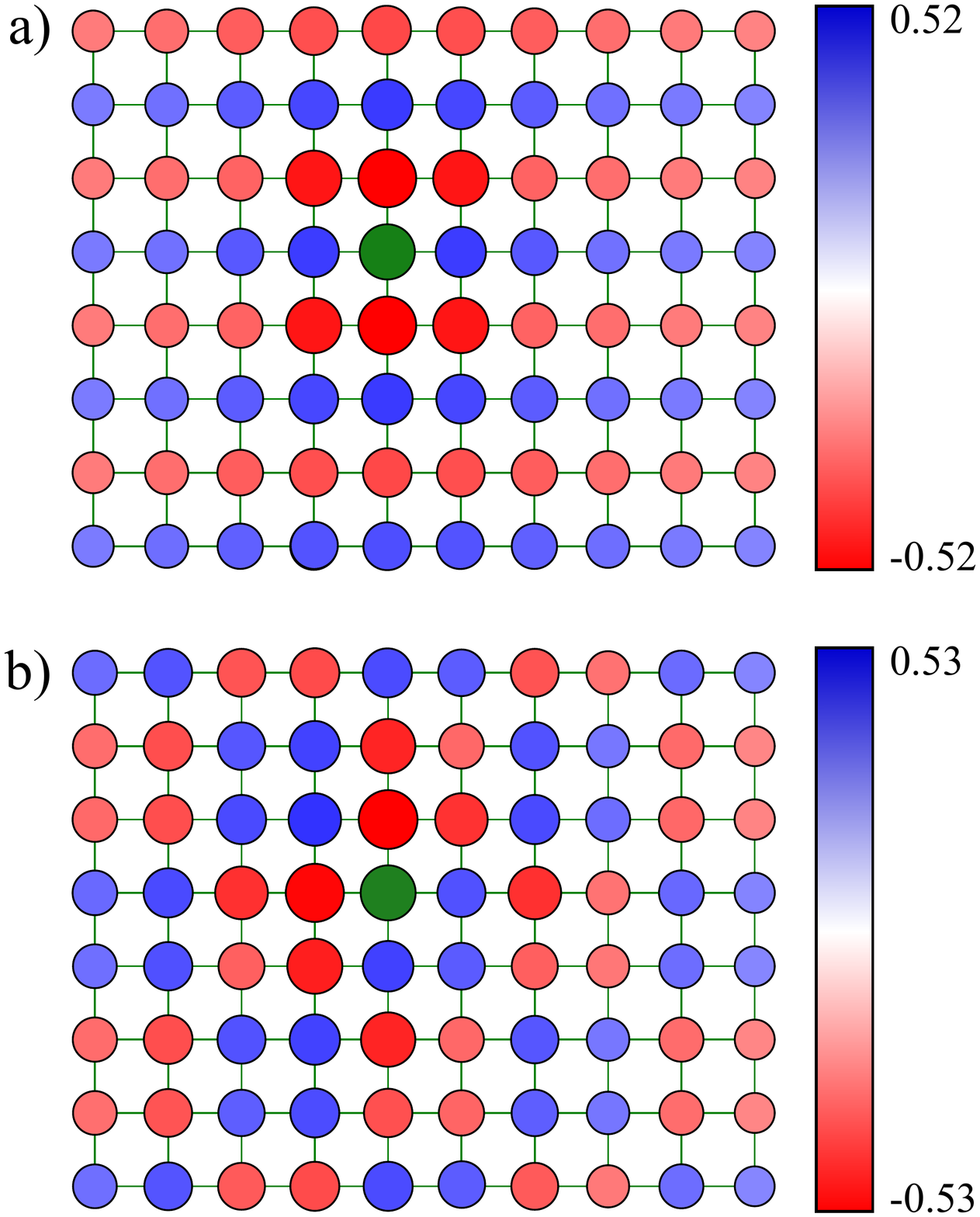}
	\end{center}
	\caption{The real-space spin correlations in the middle of the cylinders for (a) CAFM phase at $J_3=0.2$ and (b) for PL phase at $J_3=0.8$. In both cases, $J_2=0.8$ and $K_1=K_2=-0.3$. The green site is the reference site; the blue and red colors denote positive and negative correlations of the sites with the reference site, respectively. The area of each circle is proportional to the magnitude of the spin correlation and is also indicated by the circle's color for clarity.}
	\label{k03}
\end{figure}

First we show in Fig.~\ref{k03} the real-space spin configurations for the CAFM  and the PL order obtained through the calculations of the spin-spin correlation functions by DMRG on an $L = 8$ cylinder. 
 Due to the cylindrical geometry, the CAFM automatically chooses an antiparallel configuration along the $y$ direction and a parallel configuration along the $x$ direction [see Fig.~\ref{k03}(a)]. 
Note that the PL order shown in Fig.~\ref{k03}(b) is distinct from the DS order shown in Fig.~\ref{Figds}; however, the two solutions have degenerate ground-state energies given by Eq.~(\ref{eq:E_DS}). 

\begin{figure}[!tb]
\includegraphics[width=0.5\textwidth]{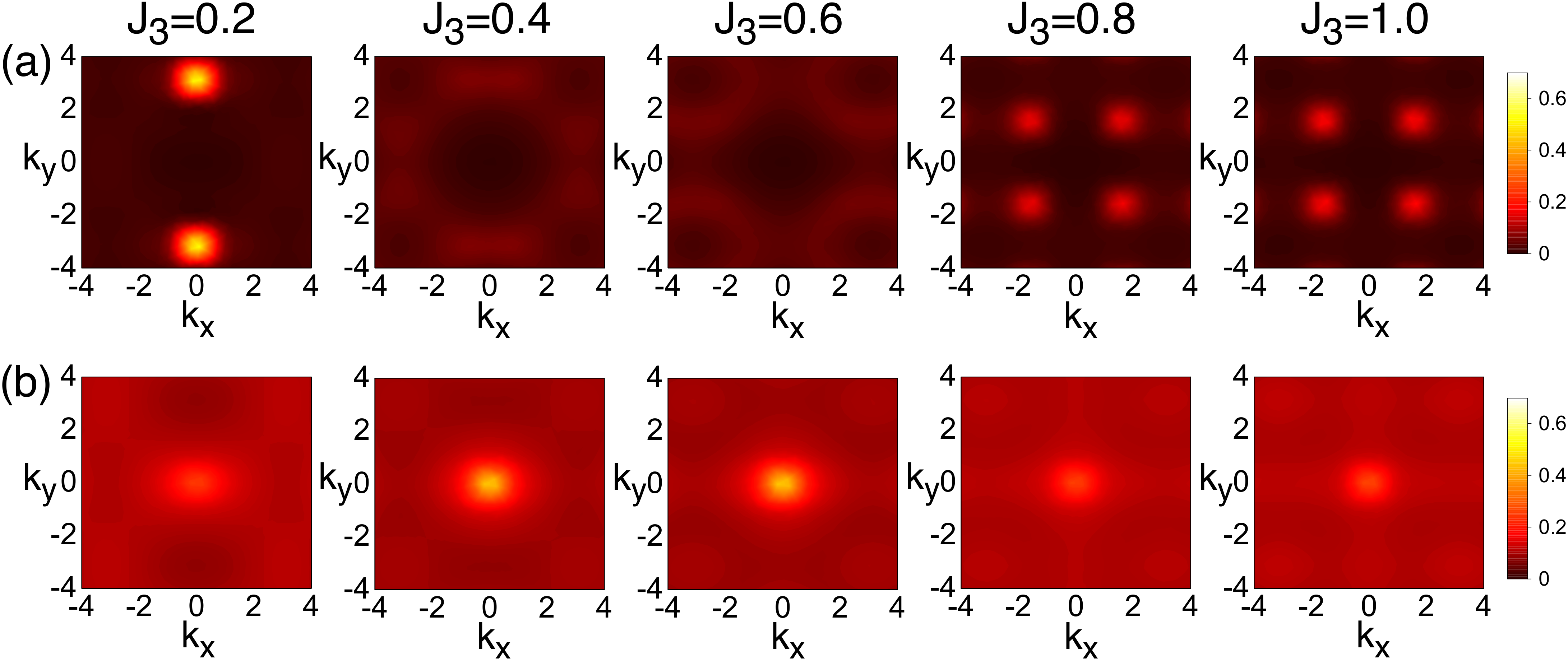}
\caption{Static spin and quadrupolar structure factors obtained from DMRG on RC$L\!\! -\!\! 2L$ cylinders with $J_1=1, J_2=0.8, K_2=K_1=-0.3$ as a function of $J_3$. 
(a) First row:  $m_S^2(\bm{q})$ for $L=8$.
(b) Second row: $m_Q^2(\bm{q})$ for $L=8$.
}
\label{Fig2}
\end{figure}

\begin{figure}[!tb]
\includegraphics[width=0.5\columnwidth]{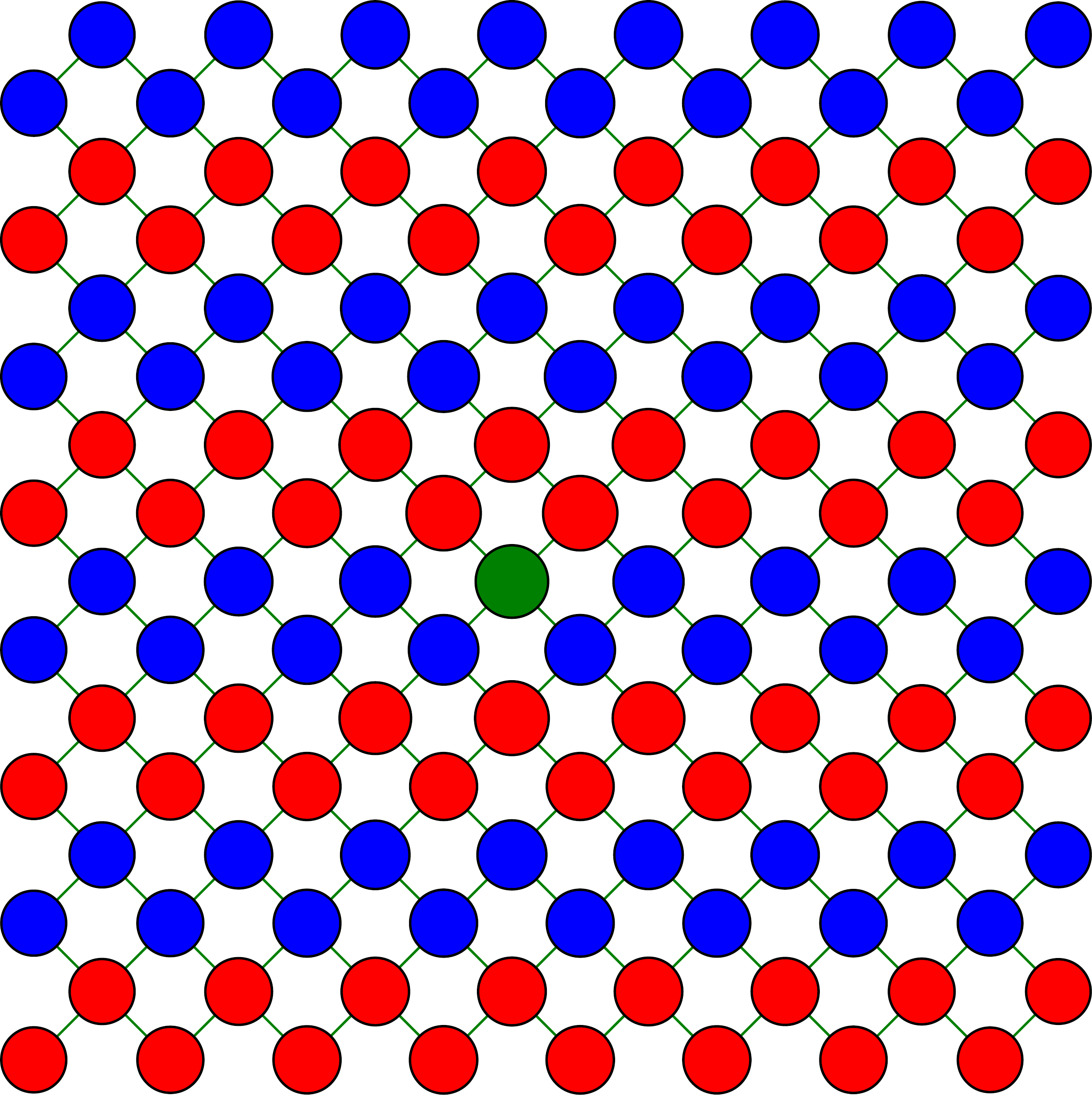}
\caption{The real-space spin correlations for DS phase at $J_3=0.8$ on the tilted geometry with $J_2=0.8$ and $K_1=K_2=-0.3$. The green site is the reference site; the blue and red colors denote positive and negative correlations of the sites with the reference site, respectively. The area of the circle is proportional to the magnitude of the spin correlation.}
\label{Figds}
\end{figure}

In order to analyze the structure of the spin correlations in different phases, it is more convenient to work in reciprocal space. Shown in Fig.~\ref{Fig2}(a) for different values of $J_3$ is the static spin-structure factor 
\beq
m^{2}_{S}(\bm{q})=\frac{1}{L^4}\sum_{ij} \langle {\bf S}_{i}\cdot {\bf S}_{j}\rangle e^{i\bm{q}\cdot(\bm{r}_i-\bm{r}_j)}.
\eeq
In the above formula, the indices $i,j$ are only partially summed on $L \times L$ sites in the middle of the cylinder in order to reduce boundary effects \cite{white2007,Yan2011,gong2014square,Gong2015_honeycomb}. 
The leftmost panel, at $J_3=0.2$, is in the CAFM phase and corresponds to the real-space spin configuration shown earlier in Fig.~\ref{k03}(a). Predictably, $m_S^2(\bm{q})$ is maximized at $\bm{Q}_1=(0,\pi)$ (as noted above, the cylindrical DMRG geometry selects $\bm{Q}_1$ over $\bm{Q}_2$). 
At $J_3 \gtrsim 0.8$, a PL phase becomes stable, with the spin-structure factor attaining a maximum at $\bm{Q}=(\pi/2,\pi/2)$.
The  $J_3=0.8$ panel in Fig.~\ref{Fig2}(a) corresponds to the real-space configuration shown in Fig.~\ref{k03}(b) above. 

In between the CAFM and the PL phase, the static spin-structure factor is featureless, indicative of the absence of the conventional static magnetic long-range order. In order to shed more light on the nature of spin correlations in this phase, we have calculated the static spin-quadrupolar structure factor, defined as
\beq
m^{2}_{Q}(\bm{q})=\frac{1}{L^4}\sum_{ij} \langle \bm{\mathit Q}_{i}\cdot \bm{\mathit Q}_{j}\rangle e^{i\bm{q}\cdot(\bm{r}_i-\bm{r}_j)}.
\eeq
The results are shown in Fig.~\ref{Fig2}(b) as a function of increasing $J_3$.
On general grounds, one expects nonzero quadrupolar correlations inside conventional long-range order phases, such as the CAFM (leftmost panel) and the PL (two rightmost panels). However, it is the  intermediate regime $0.2 \lesssim J_3 \lesssim 0.8$ that is most interesting. In this phase, $m^{2}_{Q}$ has a pronounced maximum at $q=(0,0)$, whereas the spin-structure factor is featureless in Fig.~\ref{Fig2}(a), corroborating the ferroquadrupolar nature of this phase. 


By comparing the DMRG results with the mean-field phase diagram in Fig.~\ref{Fig1}(c), we observe that the FQ phase occupies a much wider region in DMRG, whereas it is only predicted to be stable at $K_1=K_2<K_c$ ($K_c^{MF}=-2/3$) by the mean-field analysis. 
This is consistent with our earlier DMRG results at $J_3=0$ in Ref.~\cite{ZWang2016}, which also found the FQ phase to be stable in a wider region than the mean-field prediction. 

As was mentioned in Secs.~\ref{sec:MF} and~\ref{sec:phd} above, the bicollinear DS phase [Fig.~\ref{Fig:DS}(a)] and the PL phase [Fig.~\ref{Fig:DS}(b)] have the same energy within the mean-field calculation. Our DMRG calculations indicate that either of the two phases can be stabilized, depending on the setup geometry. Namely, we find the PL phase to be the ground state in the $L=8$ rectangular geometry, whereas the DS phase is favored by the tilted geometry (cylindrical axis at 45$^\circ$ angle to the lattice base vectors). The energies of the two phases at $J_3=0.8$ and $K_1=K_2=-0.3$ on the $L=8$ cylinder are very close to each other: $E_\text{rect}=-3.88345$ and $E_\text{tilt}=-3.87157$, respectively, making the DMRG inconclusive as to the choice of the ground state. It was shown recently that the apparent degeneracy of the two states is robust over a wide parameter regime and persists to higher spin values ($S=3/2, 2$); the four-site ring-exchange interaction lifts the degeneracy, favoring the DS order~\cite{Lai2016}.

\section{Dynamical spin structure factor and comparison with experiment}\label{sec:Sqw}

\begin{figure}[!tb]
\includegraphics[width=0.5\textwidth]{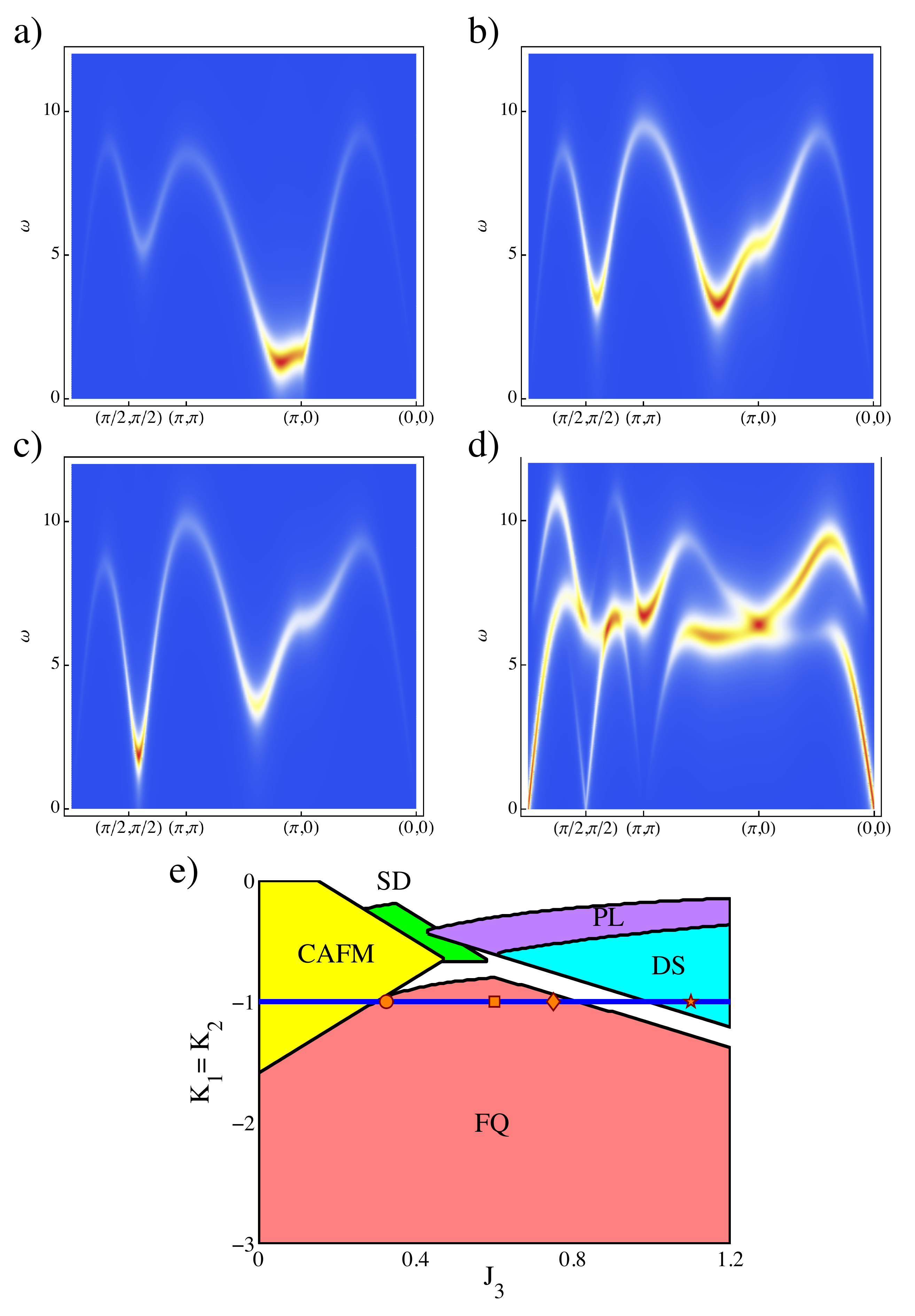}
\caption{Dynamic spin-structure factor $S(\bm{q},\omega)$, calculated along the horizontal cut through $K_1=K_2\equiv K=-1$ in Figs.~\ref{Fig1}(d) and \ref{Fig1}(h):
(a)-(c) inside FQ phase at $J_3=0.325, 0.6$, and $0.75$, respectively; (d) inside the DS phase at $J_3=1.1$. These points are indicated in the corresponding cut of the phase diagram (e) by the circle, the square, the diamond and the asterisk, respectively.
}
\label{Fig3}
\end{figure}

 Experimentally, the \FeTeSe\, series provides a unique opportunity to study the transition from the nonmagnetic FeSe to the double-stripe phase in Fe$_{1+y}$Te. Given the interpretation advanced in Ref.~\cite{ZWang2016} that FeSe has the FQ ground state, it is very interesting to study the transition from the FQ to DS phase and compare with the available INS data on \FeTeSe. Our calculations (see Fig.~\ref{Fig1}) indicate that a sizable value of $J_3$ is required in order to stabilize the DS phase observed in FeTe. It is therefore natural to mimic Te doping of FeSe with increasing the value of $J_3$.
  To this end, we have calculated the dynamic spin-structure factors $S(\vq,\omega)$ from the flavor-wave expansion (see Appendix~\ref{app:flwave}) with increasing $J_3$ along the horizontal cut along $K_1=K_2\equiv K=-1$ in Fig.~\ref{Fig1}(d). These are shown in Fig.~\ref{Fig3}. Of course we realize that other parameters will generically also be affected by Te doping, charting a complex path in the phase space of the model; however, since we are after the qualitative trend, this admittedly simplified picture is justified.

According to our phase diagrams in Fig.~\ref{Fig1}, the CAFM phase is separated from the DS phase by either the nonmagnetic FQ phase for $K_1<K_c$ or by the intermediate magnetic SD or PL phase for $K_1>K_c$.
 While it is possible to fine-tune the model parameters in such a way as to make the transition from CAFM to DS direct [see, e.g., Fig.~\ref{Fig1}(c)], the above presented scenario is generic. In Fig.~\ref{Fig3}, we have chosen such a generic cut of the phase diagram across the FQ phase, and we follow the evolution of the spin-structure factor as the DS phase is approached from inside the FQ phase.

Inside the FQ ground state, however, the spin-rotational symmetry of the Hamiltonian Eq.~(\ref{eq:model}) is broken without breaking the time-reversal symmetry. The resulting Goldstone modes at $\vq=0$ therefore have vanishing intensity ($S(0,\omega) \propto \omega$~\cite{Lauchli2006,Tsunetsugu2006}) in the static limit, consistent with the absence of Bragg peaks in FeSe under ambient pressure~\cite{McQueen2009,Bendele2010}. For small $J_3$, close to the CAFM boundary, the
 spin-structure factor has pronounced minima at $\bm{Q}_{1,2}=(\pi,0)/(0,\pi)$ as can be seen in Fig.~\ref{Fig3}(a), in accord with the INS on FeSe~\cite{Rahn2015,  QWang2015, QWang2016,Shamoto2015}. Upon increasing $J_3$, we observe another set of peaks at $\bm{Q}_{3,4}=[\pi/2+\delta,\pm (\pi/2+\delta)]$ become lower in energy [Figs.~\ref{Fig3}(b) and \ref{Fig3}(c)]. This is especially pronounced close to the boundary with the DS phase [Fig.~\ref{Fig3}(c)]. These are generally incommensurate ($\delta\neq 0$, see Fig.~\ref{Fig4}); eventually, these peaks evolve into the Goldstone modes inside the DS phase when $\delta=0$. 
These features are in qualitative accord with the INS data on \FeTeSe, where the low-energy spin excitations evolve from being dominated by the $\bm{Q}_{1,2}$ minima for $x\approx1$~\cite{Qiu2009, Lee2010, Lumsden_NPhys2010,Liu2010} to that of Fe$_{1+y}$Te~\cite{Bao2009,Li2009,Wen2009} upon decreasing $x$. 

\begin{figure}[!t]
	\includegraphics[width=0.5\textwidth]{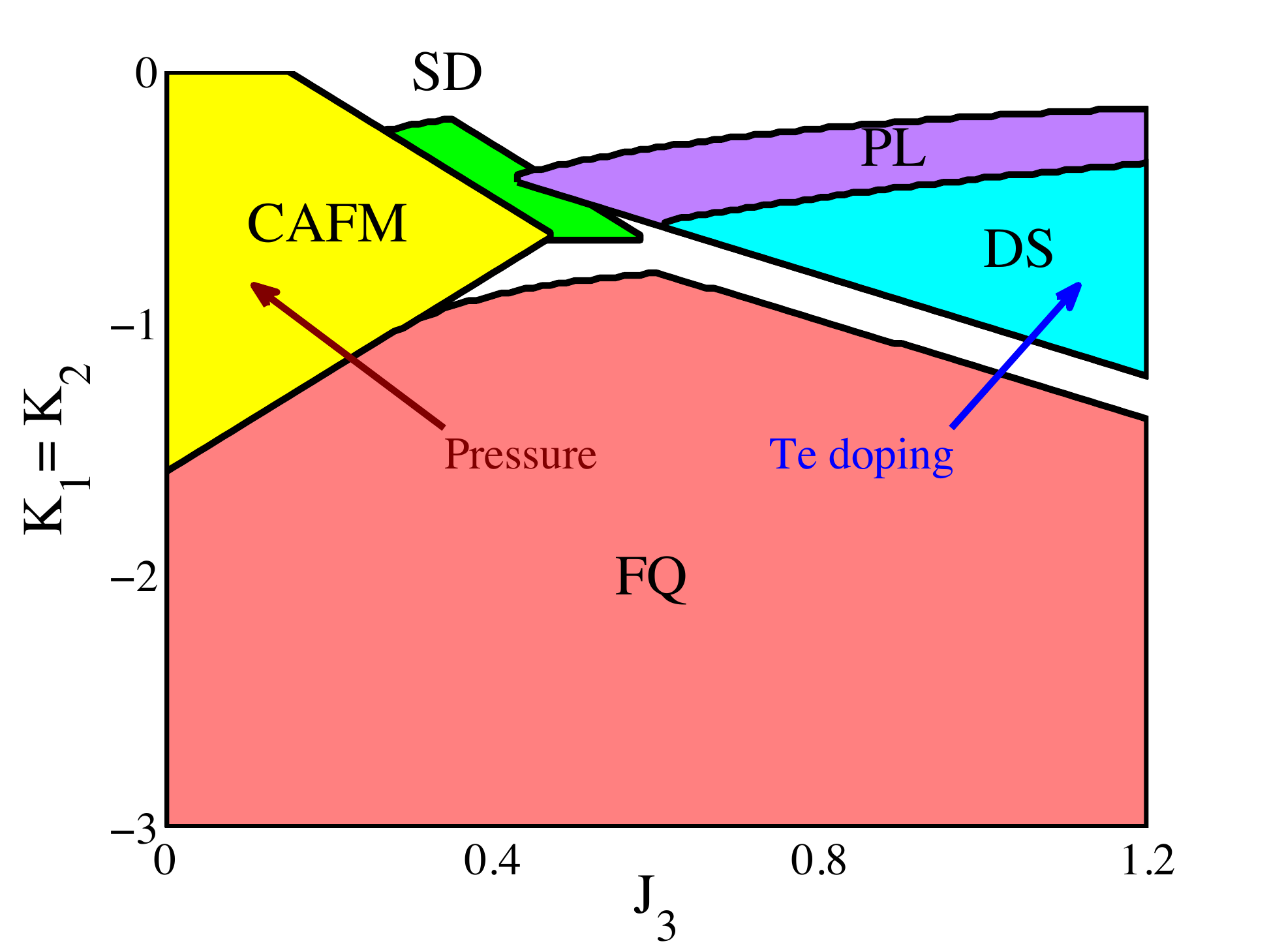}
	\caption{Trajectories in the space of the model parameters that qualitatively correspond to the transitions from the nonmagnetic phase of FeSe into various magnetically ordered states upon applied pressure and Te doping.
	}
	\label{Fig10}
\end{figure}

It has been reported that applying pressure to FeSe leads to the onset of magnetism~\cite{Bendele2012,Terashima2015,Kothapalli2016}, reportedly the CAFM phase~\cite{Wang-NMR2016}. Comparing with Fig.~\ref{Fig1}, we conclude that the effect of applying pressure  corresponds to decreasing the ratio $x=J_3/J_1$  and decreasing the biquadratic couplings $|K_i|/J_1$ in the $(J_3,K)$ phase diagram, resulting in the transition from the FQ into the CAFM phase. This conclusion is corroborated by the \textit{ab initio} calculations by Glasbrenner \textit{et al.} in Ref.~\cite{Glasbrenner2015} who find that applying pressure to FeSe indeed suppresses the ratios of both $J_3/J_1$ and $K_1/J_1$. This trend is indicated qualitatively by a corresponding arrow in the phase space of model parameters in Fig.~\ref{Fig10}. 
%
%
Doping with Te, on the other hand, can be thought of as increasing the ratio $J_3/J_1$ and possibly also $|K_i|/J_1$, as we have remarked in the beginning of this section.
Therefore, the theoretical phase diagrams in Fig.~\ref{Fig1}, together with the trends indicated by arrows in Fig.~\ref{Fig10}, capture the salient features of both  tellurium doping and of applying hydrostatic pressure to FeSe. The actual trajectories in the phase space of the model parameters that correspond to these experimental knobs are likely more complicated; nevertheless, our analysis provides an important qualitative insight into the physics of the spin degrees of freedom in FeSe and \FeTeSe.

Intriguingly, the neutron spin structure in superconducting \FeTeSe\, samples undergoes a complicated transformation as a function of temperature, with the high-temperature data ($T\gtrsim 100$~K) characterized by the DS wave vector $(\pi/2,\pi/2)$ and evolving to $\bm{Q}_{1,2}$ upon cooling~\cite{Xu2016}. Remarkably, it was found that in nonsuperconducting \FeTeSe\, samples (due to excess of Fe), the magnetic correlations remain pinned at $(\pi/2,\pi/2)$~\cite{Xu2016}. The authors of Ref.~\cite{Xu2016} have concluded that the observed thermal change in characteristic wave vector is therefore correlated with the tendency towards nematic $xz/yz$ orbital splitting at low temperature, which favors superconductivity. Theoretical verification of these conclusions requires taking into consideration the multiorbital character of conduction electrons and is beyond the effective spin model studied in this paper. It was suggested~\cite{Moon2010} that the orbital and superexchange physics is particularly sensitive to the Fe--(Se,Te) -- Fe bond angle, which in \FeTeSe\, is controlled by the height of the chalcogenide ions above and below the Fe layers~\cite{McQueen2009b, Louca2010}. This complexity notwithstanding, we can nevertheless conclude that in the samples with excess Fe, where the aforementioned orbital effects are less pronounced, our effective spin model correctly predicts the characteristic wave vector of low-energy spin excitations to evolve from $(\pi,0)/(0,\pi)$ towards $(\pi/2,\pi/2)$ upon Te doping.

\section{Incommensurate phases}\label{sec:IC}
\begin{figure}[!tb]
\includegraphics[width=0.5\textwidth]{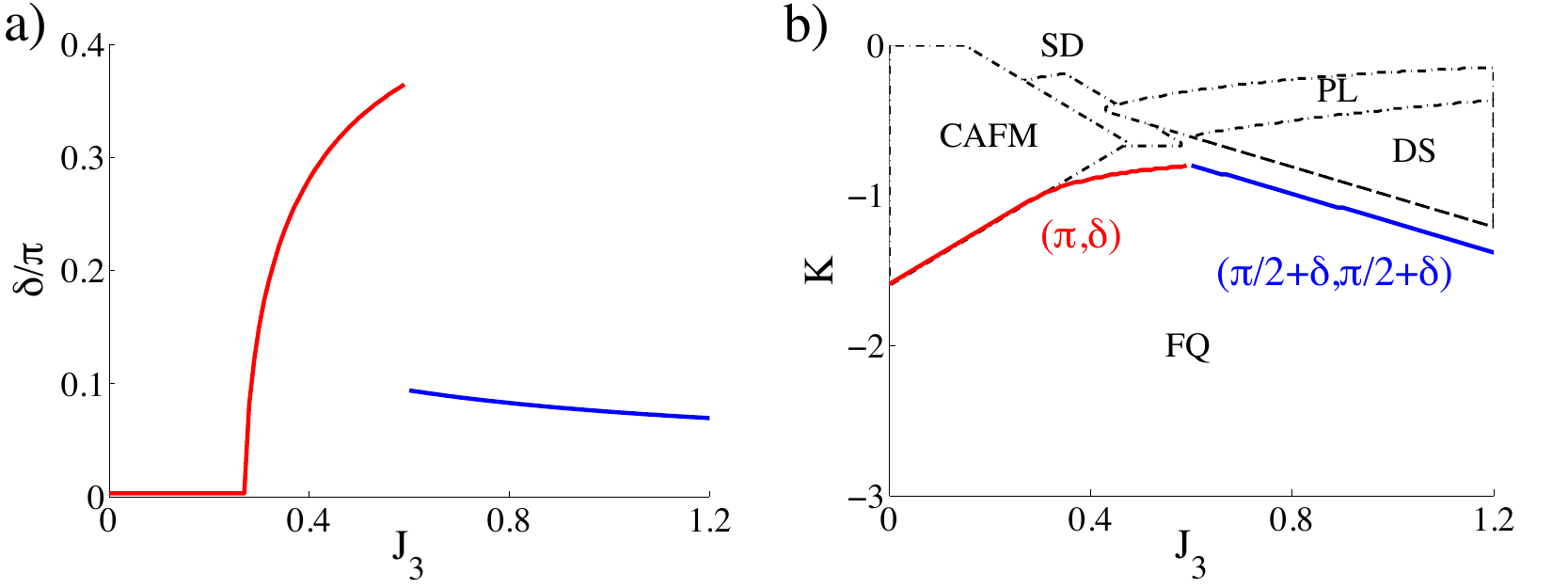}
\caption{(a) The characteristic wave vector of the flavor-wave instability along the FQ phase boundary $K=K_c(J_3)$, shown in panel (b) as a red/blue line. The red segment indicates the instability towards the $(\pi,\delta)$ phase, and the blue segment towards the $(\pi/2+\delta,\pi/2+\delta)$ phase. The remainder of the phase diagram is the same as in Fig.~\ref{Fig1}(h), calculated within the flavor-wave method as a function of $J_3$ and $K_1=K_2\equiv K$, using $J_1=1$, $J_2=0.8$. 
}
\label{Fig4}
\end{figure}

It is interesting to note that early INS experiments have indicated that the high-temperature spin-structure factor in \FeTeSe\, may be incommensurate~\cite{Xu2012, Tsyrulin2012, Xu2014}. While the latest INS data indicate that this may not in fact be the case~\cite{Xu2016}, it is instructive to consider predictions of our theory in this regard. The variational mean-field phase diagrams in Figs.~\ref{Fig1}(a)-\ref{Fig1}(d) contain only commensurate phases, which is understandable given the constraint that the solution must obey the periodic boundary conditions on a $4\times4$ Fe-site cluster. Similarly, the DMRG on cylindrical geometry is limited to small $L\leq 8$, which makes the search for an incommensurate spiral phase very difficult. The flavor-wave analysis, on the other hand, is by its nature a thermodynamic expansion around the mean-field solution and is not limited to commensurate wave vectors. As noted earlier, the white regions in Figs.~\ref{Fig1}(e)-\ref{Fig1}(h) indicate an instability of the flavor-wave expansion towards other solutions, including incommensurate spin spiral states. In order to shed more light on the issue, we have analyzed the low-energy dynamical spin-structure factor near the FQ phase boundaries $K=K_c(J_3)$ approaching the unstable white regions. In this regime, we  find softening of the flavor-wave dispersion at certain (generally incommensurate) wave vectors, which indicates a tendency towards respective magnetic ordering. The wave vectors of these unstable modes are shown in Fig.~\ref{Fig4}. 

We see from Fig.~\ref{Fig4} that for small $J_3$ near the CAFM boundary, the flavor-wave instability is pinned at $(\pi,0)/(0,\pi)$, as reported in Ref.~\cite{ZWang2016}. Upon increasing $J_3$, the characteristic wave vector becomes incommensurate $(\pi,\delta)$, with $\delta$ increasing smoothly towards, but stopping shy of $\pi/2$. 
 For higher $J_3$, upon approaching the DS phase boundary from inside the FQ phase [blue line in Fig.~\ref{Fig4}(b)] , the flavor-wave dispersion softens at an incommensurate $(\pi/2+\delta,\pi/2+\delta)$ wave vector.
While true long-range incommensurate order cannot be studied in this way for technical reasons (flavor-wave expansion around commensurate $\bm{Q}$ becomes unstable), the above analysis provides a strong indication that the reported soft modes would eventually become  Bragg peaks as the incommensurate long-range order settles in.
  
\section{Discussion}\label{sec:conclusions}

In this work, we have advanced an effective spin model for iron chalcogenides in an effort to better understand the evolution of the neutron-scattering spectra in FeSe upon applying pressure and tellurium doping. Our starting point is the strong-coupling approach, justified in the limit when Coulomb interaction $U$ is considerably larger than the electron hopping  $t$. Although the iron chalcogenides are not charge insulating systems, the strong-coupling approaches have been successfully used to elucidate many aspects of these materials, from the nature of electron nematicity~\cite{Fang2008,Yu2012,Ergueta2015} and effects of orbital selectivity~\cite{Yu2012-OSMP,Bascones2012,Yu2013,Medici2014}, to the origin of the superconducting pairing~\cite{Seo2008,Chen2009,Goswami2010,Yu2013-SC,Yu2014}. One of the justifications for using the strong-coupling approach is the large fluctuating iron moment observed in inelastic neutron scattering ($M_{eff}^2\sim 5 \mu_B^2$ per Fe ion~\cite{Wen2015}), which is difficult to obtain in the weak-coupling scenario from considering only the electrons near the Fermi surface.
 This is not to say that the conduction electrons are somehow unimportant -- on the contrary, they are crucial for superconductivity and the multiorbital effects that are beyond the scope of this work. Nevertheless, we  argue that the presented effective spin model is important for understanding the effects of magnetic frustrations in the iron chalcogenides, which have been brought to focus most prominently by the surprising observation of the apparently nonmagnetic ground state in FeSe~\cite{McQueen2009,Bendele2010,Rahn2015,  QWang2015, QWang2016,Shamoto2015}. Having proposed an explanation for this state in terms of the spin quadrupolar order in an earlier work~\cite{ZWang2016}, the present study seeks to provide an accurate, consistent description of the spin degrees of freedom in both the iron pnictides and chalcogenides within a single  microscopic spin model. Although constructing such an effective model inevitably required simplifications of the multiorbital nature of these materials, the agreement that we have obtained with the INS experiments testifies to the importance of utilizing this effective description.

In summary, we have demonstrated that the evolution of the low-energy spin excitations in FeSe under applied pressure and tellurium doping in \FeTeSe\, can be successfully understood within a single effective spin model. Although the exact dependence of the model parameters on these experimental variables is unknown, we sketch in Fig.~\ref{Fig10} the approximate corresponding trajectories in the model phase space, based on the analysis of our computed phase diagrams and consistent with prior \textit{ab initio} calculations~\cite{Glasbrenner2015}. Using a combination of analytical techniques and state-of-the-art DMRG calculations, we have established the phase diagram of the effective model and computed the dynamical spin response. In particular, the calculated evolution of the characteristic wave vector of the spin excitations matches that observed in INS experiments on \FeTeSe\, and the possibility of the incommensurate spin orders upon Te doping has been analyzed in detail.
The effects of conduction electrons, while of course very important, are beyond the scope of this effective spin model; nevertheless, given the recently observed correlation between superconductivity in \FeTeSe\, and the appearance of the $(\pi,0)/(0,\pi)$ inelastic peaks in the low-temperature dynamical spin correlation~\cite{Xu2016} puts the present work in a wider context of superconductivity in iron chalcogenides. This connection certainly deserves further theoretical study, perhaps within the framework of realistic multiorbital models that should take into account the essential features predicted by the effective spin model presented here.

\begin{acknowledgements}
We would like to thank Shou-Shu Gong for helpful discussions and providing access to the DMRG code.  We would also like to acknowledge the valuable help from Zhentao Wang, who participated in the initial stages of this project. P.~B.~E., W.~H. and A.~H.~N. acknowledge the support of NSF CAREER Grant No. DMR-1350237. A.~H.~N. was also supported by the Welch Foundation through Grant No. C-1818. W.~H. also acknowledges support from NSF Grant No. DMR-1309531. P.~B.~E. was partially funded by the Grants No. 22799 and No. 23341 from the Research Corporation for Science Advancement.
Computational resources were provided by the Big-Data Private-Cloud Research Cyberinfrastructure MRI Award funded by the NSF under Grant No. CNS-1338099 and by Rice
University, and by the Extreme Science and Engineering Discovery Environment (XSEDE)\cite{Towns2014}, which is supported by NSF Grant No. OAC-1053575.

 
\end{acknowledgements}



\appendix

\section{Flavor Wave Calculation of Dynamical Spin Structure Factors}\label{app:flwave}
%
%
%
%

By virtue of the fluctuation--dissipation theorem, the dynamical spin structure factor at $T=0$, $S^{\alpha\beta}(\qq,\omega)$ is proportional to the imaginary part of the spin susceptibility:
\begin{align} \label{Eq:sqw}
	&\quad S^{\mu\nu}(\qq,\omega)= \chi_{\mu \nu}^{\prime \prime} (\qq,\omega) \nonumber \\
	&=\!\! \frac{N_S}{N} \sum_{\alpha \beta} \sum_f \langle \gs | S^\mu_{\alpha, \bm{q}} |f\rangle \langle f|S^\nu_{\beta,-\bm{q}} |\gs\rangle \delta(\omega-E_f+E_g),
\end{align}
where $| f \rangle \langle f| = 1$ is the complete set of states, $\{ \alpha,\beta \}$ denote different sublattices, and $N/N_S$ is the total number of different sublattices. 

For magnetically ordered states, the ground state will add nonzero elastic contribution $\sim \delta(\omega)$ to $\chi_{\mu \nu}^{\prime \prime} (\qq,\omega)$, as shown in the following subsections.
For the ferroquadrupolar state, on the other hand, the ground state $|f \rangle = | g.s. \rangle$ does not contribute to $\chi_{\mu \nu}^{\prime \prime} (\qq,\omega)$ and consequently, no magnetic Bragg peak is found at $\omega=0$ in elastic neutron scattering. This can be readily understood since the quadrupolar states do not break time-reversal symmetry and as a result, do not couple in the static limit to the neutron spin. 


\subsection{Flavor Wave for FQ}
In the FQ state the directors $\vec{d}_i$ are identical on all sites (in total one sublattice $N/N_S=1$). Due to the spontaneous breaking of the  SU$(2)$ symmetry in the FQ phase, we can conveniently choose the director  corresponding to the quadrupolar order parameter to lie along the ${z}$-direction:
\begin{equation}
	\vec{d}_i=\{ 1, \, 0, \, 0 \}.
\end{equation}

\noindent
Correspondingly, the transformation matrix $\mathcal{V}_i$ defined in Eq.~(\ref{Eq.V-matrix}) is simply an identity matrix and is the same on every site $i$:
\begin{equation}
	\mathcal{V}_i =
	\begin{pmatrix}
		1 & 0 & 0 \\
		0 & 1 & 0 \\
		0 & 0 & 1
	\end{pmatrix}.
\end{equation}

\noindent
The local constraint on the condensed boson number, 
\begin{equation}
\tilde{b}_{i,0}=\sqrt{1-\tilde{b}_{i,1}^\dagger \tilde{b}_{i,1}-\tilde{b}_{i,2}^\dagger \tilde{b}_{i,2}},
\end{equation}
can be expanded up to quadratic terms in the  boson creation/annihilation operators, resulting in:
\begin{equation}
	\begin{aligned}
	\mathcal{H}_\text{fw} \!\! &=\!\! \sum_{\bm{q},a} \left[t(\bm{q})+\lambda \right] (\tilde{b}_{\bm{q},a} \tilde{b}_{\bm{q},a}^{\dagger}+\tilde{b}_{\bm{-q},a}^\dagger \tilde{b}_{-\bm{q},a}) +\\
	&+\sum_{\bm{q},a} \left[ \Delta(\bm{q}) \tilde{b}_{\bm{q},a}^\dagger \tilde{b}_{-\bm{q},a}^\dagger  +H.c. \right]+N E_0,
	\end{aligned}
\end{equation}
where:
\begin{subequations}
	\begin{align}
		t(\bm{q}) &= J_1 (\cos q_x + \cos q_y) + 2J_2 \cos q_x \cos q_y \nonumber \\
		&\quad +J_3 (\cos 2q_x + \cos 2q_y), \\
		\Delta(\bm{q}) &= (K_1-J_1) (\cos q_x + \cos q_y) +2 (K_2-J_2) \cos q_x \cos q_y \nonumber \\
		&\quad -J_3 (\cos 2q_x + \cos 2q_y), \\
		\lambda &= -2(K_1+K_2) , \\
		E_0 &=4(K_1+K_2).
	\end{align}
\end{subequations}

Bogoliubov transformation:
\begin{equation}
	\alpha_{\bm{q},a} = \cosh \theta_{\bm{q}} \tilde{b}_{\bm{q},a}-\sinh \theta_{\bm{q}} \tilde{b}_{-\bm{q},a}^\dagger,
\end{equation}
with
\begin{equation}
	\tanh 2\theta_{\bm{q}} = -\frac{\Delta{(\bm{q})}}{t(\bm{q})+\lambda}.
	\label{eq:Bogoliubov}
\end{equation}

The diagonalized Hamiltonian:
\begin{equation}
	\mathcal{H}_\text{fw}=\sum_{a=1,2}\sum_{\bm{q}}\omega_{\bm{q}} (\alpha_{\bm{q},a}^\dagger \alpha_{\bm{q},a} +1/2) + N (E_0-2\lambda),
\end{equation}
where the dispersion $\omega_{\bm{q}}$ is given by:
\begin{equation}
	\omega_{\bm{q}} =2\sqrt{[t(\bm{q})+\lambda]^2-\Delta^2(\bm{q})}.
\end{equation}

Since there is only one sublattice, we can omit the sublattice indices $\{ \alpha,\beta \}$, and only use notation $S_{\pm \bm{q}}^\mu$ for the Fourier components in this subsection. To calculate the dynamic spin susceptibility, the spin operators in Eq.~\eqref{Eq:sqw} are kept up to linear order:
\begin{subequations}
	\begin{align}
		S^x_{\bm{q}} &= 0, \\
		S^y_{\bm{q}} &= -i \left( \tilde{b}_{-\bm{q},2}^\dagger -\tilde{b}_{\bm{q},2} \right), \\
		S^z_{\bm{q}} &= i  \left( \tilde{b}_{-\bm{q},1}^\dagger -\tilde{b}_{\bm{q},1}  \right).
	\end{align}
\end{subequations}

Then Eq.~\eqref{Eq:sqw} can be written down explicitly:
\begin{subequations}
	\begin{align}
		\chi_{xx}^{\prime \prime} (\bm{q},\omega) &= 0, \\
		\chi_{yy}^{\prime \prime} (\bm{q},\omega) &= \frac{t(\bm{q}) + \lambda +  \Delta(\bm{q})}{\sqrt{[t(\bm{q})+\lambda]^2-\Delta^2(\bm{q})}} \delta(\omega  - \omega_{\bm{q}}), \label{chi_yy}\\
		\chi_{zz}^{\prime \prime} (\bm{q},\omega) &= \frac{t(\bm{q}) + \lambda +  \Delta(\bm{q})}{\sqrt{[t(\bm{q})+\lambda]^2-\Delta^2(\bm{q})}} \delta(\omega - \omega_{\bm{q}}). \label{chi_zz}
	\end{align}
\end{subequations}

Note that at $\omega_{\bm{q}}=0$, the Bogoliubov angle $\theta_{\bm{q}}=0$ in Eq.~(\ref{eq:Bogoliubov}) and it follows that $t(\bm{q}) + \lambda +  \Delta(\bm{q}) = 0$ in the numerator on Eqs.~(\ref{chi_yy}) and (\ref{chi_zz}). We see that as a result, the spin structure factor vanishes at $\bm{q}=0$, in other words, the Goldstone mode of the FQ state does not contribute to the static spin susceptibility, as seen in Fig.~\ref{Fig3}. This fact is well known for the quadrupolar states~\cite{Lauchli2006,Tsunetsugu2006,ZWang2016}) and is consistent with the absence of the magnetic Bragg peaks in the elastic neutron scattering in FeSe~\cite{McQueen2009,Bendele2010}. 

\subsection{Flavor Wave for CAFM}
There are in total two sublattices $N/N_S=2$, whose directors can be chosen as:
\begin{subequations}
	\begin{align}
		\vec{d}_{i \in A} &=\frac{1}{\sqrt{2}}\{0,1,i\}, \\
		\vec{d}_{i \in B} &=\frac{1}{\sqrt{2}}\{0,1,-i\}.
	\end{align}
\end{subequations}

Correspondingly, the transformation matrices are written below:
\begin{subequations}
	\begin{align}
		\mathcal{V}_{i \in A} &=\frac{1}{\sqrt{2}}
		\begin{pmatrix}
			0 & 0 & \sqrt{2} \\
			1 & i & 0 \\
			i & 1 & 0
		\end{pmatrix}, \\
		\mathcal{V}_{i \in B} &=\frac{1}{\sqrt{2}}
		\begin{pmatrix}
			0 & 0 & \sqrt{2} \\
			1 & -i & 0 \\
			-i & 1 & 0
		\end{pmatrix}.
	\end{align}
\end{subequations}

The quadratic terms of the resulting Hamiltonian now include cross terms between sublattices:

\begin{equation}
	\begin{aligned}
		\mathcal{H}_\text{fw} \!\! &=\!\! \sum_{\bm{q},a} \left( t_{aa}+\lambda_{aa} \right) (\tilde{b}_{A,\bm{q},a} \tilde{b}_{A,\bm{q},a}^{\dagger}+\tilde{b}_{A,\bm{q},a}^\dagger \tilde{b}_{A,\bm{q},a} +\nonumber \\
		&\qquad\qquad\quad\quad +\tilde{b}_{B,\bm{q},a} \tilde{b}_{B,\bm{q},a}^{\dagger}+\tilde{b}_{B,\bm{q},a}^\dagger \tilde{b}_{B,\bm{q},a} ) +\\
		&\quad +\sum_{\bm{q},a}  \Delta_{aa}\left(\tilde{b}_{A,\bm{q},a}^\dagger \tilde{b}_{B,-\bm{q},a}^\dagger+\tilde{b}_{B,\bm{q},a}^\dagger \tilde{b}_{A,-\bm{q},a}^\dagger  +h.c. \right)+NE_0.
	\end{aligned}
\end{equation}

With the coefficients $\lambda_{aa}$, $t_{aa}(\vq)$ and $\Delta_{aa}(\vq)$ depending on the parameters of the model as follows:

\begin{subequations}
	\begin{align}
		\lambda_{11}=&2(2J_2-K_2)-4J_3, \\
		\lambda_{22}=&-K_1+2(J_2-K_2)-2J_3,\\
		t_{11}(\vq)=&K_1\cos{q_y}, \\
		t_{22}(\vq)=&J_1\cos{q_y}+J_3[\cos{(2q_x)}+\cos{(2q_y)}]\\
		\Delta_{11}(\vq)=&K_1\cos{q_x}+2K_2\cos{q_x}\cos{q_y}, \\
		\Delta_{22}(\vq)=&-(J_1-K_1)\cos{q_x}-\nonumber \\
		&-2(J_2-K_2)\cos{q_x}\cos{q_y},\\
		E_0=&3K_1-2J_2+4K_2+2J_3.
	\end{align}
\end{subequations}

The diagonalized Hamiltonian looks as follows:

\begin{equation}
\mathcal{H}_\text{fw}=\sum_{a=1,2}\sum_{\bm{q}}\omega_{\bm{q},a} (\alpha_{\bm{q},a}^\dagger \alpha_{\bm{q},a} + \beta_{\bm{q},a}^\dagger \beta_{\bm{q},a} + 1) + N (E_0-\lambda_{11}-\lambda_{22}),
\end{equation}

and the diagonalized Bogolibouv dispersions finally take the following form (with $a=1,2$):

\begin{equation}
	\omega_{\vq,a} =2\sqrt{[t_{aa}(\vq)+\lambda_{aa}]^2-\Delta_{aa}^2(\vq)}.
\end{equation}

\subsection{Flavor Wave for N\'eel State}

In this case, both the Hamiltonian as well as the diagonalized dispersions have the same symbolic expression as in the CAFM case. However, the coefficients are now given by

\begin{subequations}
	\begin{align}
		\lambda_{11}=&2(2J_1-K_1)-2(2J_2-K_2)-4J_3, \\
		\lambda_{22}=&2(J_1-K_1)-2J_2-2J_3,\\
		t_{11}(\vq)=&2K_2\cos{q_x}\cos{q_y}, \\
		t_{22}(\vq)=&2J_2\cos{q_x}\cos{q_y}\nonumber \\
		&+J_3[\cos{(2q_x)}+\cos{(2q_y)}],\\
		\Delta_{11}(\vq)=&K_1(\cos{q_x}+\cos{q_y}), \\
		\Delta_{22}(\vq)=&-(J_1-K_1)(\cos{q_x}+\cos{q_y}),\\
		E_0=&-2J_1+4K_1+2J_2+2K_2+2J_3.
	\end{align}
\end{subequations}

\subsection{Flavor Wave for DS}

Unlike in the previous two cases where the introduction of two sublattices was enough, four are necessary in this case. However, since there are still only two distinct directors, the previously shown transformation matrices are enough to derive the Hamiltonian. It is now convenient to write the actual Hamiltonian down so that it becomes block diagonal. This is due to the lack of cross terms between the bosonic operators of the different modes. The quadratic terms can be written in the following matricial form,

\begin{equation}
	\mathcal{H}_\text{fw} \!\! =\!\! 2\sum_{\bm{q}} (\psi^{\dagger}_{11} \psi_{22}^{\dagger})H_{\text{fw}}
	\begin{pmatrix}
		\psi_{11}\\
		\psi_{22} 
	\end{pmatrix}+NE_0,
\end{equation}
with the block-diagonal form of the Hamiltonian matrix explicitly written as

\begin{equation}
\tilde{S}_i^\nu= \mathcal{V}_i^\dagger S_i^\nu \mathcal{V}_i,
\end{equation}

\begin{equation}
H_{\text{fw}}=\begin{pmatrix}
\mathcal{J} & 0 \\
0 & \mathcal{K}
\end{pmatrix},
\label{eq.matrix}
\end{equation}
and where $\psi_{aa}=(b_{A,\bm{q},a},b_{B,\bm{q},a},b_{C,\bm{-q},a}^{\dagger},b_{D,\bm{-q},a}^{\dagger})^T$. The matrix elements of each $4\times4$ block-diagonal matrix are given by

\begin{subequations}
	\begin{align}
		\mathcal{J}_{11}&=\mathcal{J}_{22}=\mathcal{J}_{33}=\mathcal{J}_{44}\nonumber \\
		&=4J_3+K_2\cos{(q_x-q_y)}\equiv\mathcal{A}, \\
		\mathcal{J}_{12}&=\mathcal{J}_{14}^*=\mathcal{J}_{21}^*=\mathcal{J}_{23}=\mathcal{J}_{32}^*=\mathcal{J}_{34}=\mathcal{J}_{41}=\mathcal{J}_{43}^*\nonumber \\
 		& = \frac{K_1}{2}(e^{iq_x}+e^{iq_y}), \\
		\mathcal{J}_{13}&=\mathcal{J}_{24}=\mathcal{J}_{31}=\mathcal{J}_{42}=K_2\cos{(q_x+q_y)}\equiv\mathcal{B},
	\end{align}
\end{subequations}
and
\begin{subequations}
	\begin{align}
		\mathcal{K}_{11}&=\mathcal{K}_{22}=\mathcal{K}_{33}=\mathcal{K}_{44}\nonumber \\
		&=-(K_1+K_2)+2J_3+J_2\cos{(q_x-q_y)}\equiv\mathcal{C}, \\
		\mathcal{K}_{12}&=\mathcal{K}_{21}^*=\mathcal{K}_{34}=\mathcal{K}_{43}^*=\frac{J_1}{2}(e^{iq_x}+e^{iq_y}), \\
		\mathcal{K}_{14}^*&=\mathcal{K}_{23}=\mathcal{K}_{32}^*=\mathcal{K}_{41}=-\frac{(J_1-K_1)}{2}(e^{iq_x}+e^{iq_y}), \\
		\mathcal{K}_{13}&=\mathcal{K}_{24}=\mathcal{K}_{31}=\mathcal{K}_{42}=-(J_2-K_2)\cos{(q_x+q_y)}\nonumber\\
		&-J_3[\cos{(2q_x)}+\cos{(2q_y)}]\equiv\mathcal{D}.
	\end{align}
\end{subequations}

And the constant terms of the energy are $E_0=3K_1+3K_2-2J_3$.

The dispersions can be derived immediately from a standard Bogoliubov transformation of the Hamiltonian above. This is done by obtaining the eigenvalues of the new matrix resulting from the similarity transformation $\widetilde{H}_{\text{fw}}=\Theta H_{\text{fw}}$, where the matrix $\Theta=\text{diag}(1,1,-1,-1)$. This gives the following result:

\begin{subequations}
	\begin{align}
		\omega_{\vq,1,\pm} &=2\sqrt{\mathcal{A}^2-\mathcal{B}^2\pm2\sqrt{\kappa_1}},\\
		\omega_{\vq,2,\pm} &=2\sqrt{\mathcal{C}^2-\mathcal{D}^2-\frac{K_1}{2}(K_1-2J_1)\pm2\sqrt{\kappa_2}},
	\end{align}
\end{subequations}

where $\kappa_1$ and $\kappa_2$ are given by

\begin{equation}
	\begin{aligned}
		\kappa_1=&\frac{K_1^2}{2}(\mathcal{A}^2+\mathcal{B}^2)[1+\cos{(q_x-q_y)}]\\
		&-\frac{K_1^2\mathcal{A}\mathcal{B}}{2}[\cos{(2q_x)}+\cos{(2q_y)}+2\cos{(q_x+q_y)}]\\
		&-\frac{K_1^4}{16}[\sin{(2q_x)}+\sin{(2q_y)}+2\sin{(q_x+q_y)}]^2,
	\end{aligned}
\end{equation}

\begin{equation}
	\begin{aligned}
		\kappa_2=&\frac{1}{2}[J_1^2\mathcal{C}^2+(J_1-K_1)^2\mathcal{D}^2][1+\cos{(q_x-q_y)}]\\
		&+\frac{J_1(J_1-K_1)\mathcal{C}\mathcal{D}}{2}[\cos{(2q_x)}+\cos{(2q_y)}+2\cos{(q_x+q_y)}]\\
		&-\frac{J_1^2(J_1-K_1)^2}{16}[\sin{(2q_x)}+\sin{(2q_y)}+2\sin{(q_x+q_y)}]^2,
	\end{aligned}
\end{equation}

and the diagonalized Hamiltonian is written as

\begin{equation}
\begin{aligned}
\mathcal{H}_\text{fw}&=\sum_{\sigma=\pm}\sum_{a=1,2}\sum_{\bm{q}}\omega_{\bm{q},a,\sigma} (\alpha_{\bm{q},a,\sigma}^\dagger \alpha_{\bm{q},a,\sigma} + \beta_{\bm{q},a,\sigma}^\dagger \beta_{\bm{q},a,\sigma} + 1)\\
&+ N (E_0+K_1+K_2-6J_3).
\end{aligned}
\end{equation}

In order to obtain the dynamical spin-spin structure factor, we first obtain the expressions for the spin operators. These can be immediately deduced from the rotated matrices. These are explicitly given in this case by

\begin{equation}
	\begin{aligned}
		\tilde{S}_{i\in A,B}^x&= \mathcal{V}_{i\in A,B}^\dagger S_{i\in A,B}^x \mathcal{V}_{\in A,B}=\\
		&=\frac{1}{2}\begin{pmatrix}
			0 & 1 & -i \\
			0 & -i & 1 \\
			\sqrt{2} & 0 & 0
		\end{pmatrix}
		\begin{pmatrix}
			0 & 0 & 0 \\
			0 & 0 & -i \\
			0 & i & 0
		\end{pmatrix}
		\begin{pmatrix}
			0 & 0 & \sqrt{2} \\
			1 & i & 0 \\
			i & 1 & 0
		\end{pmatrix}=
		\begin{pmatrix}
			1 & 0 & 0 \\
			0 & -1 & 0 \\
			0 & 0 & 0
		\end{pmatrix},
		\end{aligned}
\end{equation}

and similarly,

\begin{equation}
\tilde{S}_{i\in A,B}^y= \mathcal{V}_{i\in A,B}^\dagger S_{i\in A,B}^y \mathcal{V}_{\in A,B}=\frac{1}{\sqrt{2}}
		\begin{pmatrix}
			0 & 0 & -1 \\
			0 & 0 & -i \\
			-1 & i & 0
		\end{pmatrix},
\end{equation}

\begin{equation}
\tilde{S}_{i\in A,B}^z= \mathcal{V}_{i\in A,B}^\dagger S_{i\in A,B}^z \mathcal{V}_{\in A,B}=\frac{1}{\sqrt{2}}
		\begin{pmatrix}
			0 & 0 & i \\
			0 & 0 & 1 \\
			-i & 1 & 0
		\end{pmatrix}.
\end{equation}

For the remaining two sublattices, the rotated spin matrices are now

\begin{equation}
\tilde{S}_{i\in C,D}^x= \mathcal{V}_{i\in C,D}^\dagger S_{i\in C,D}^x \mathcal{V}_{\in C,D}=
		\begin{pmatrix}
			-1 & 0 & 0 \\
			0 & 1 & 0 \\
			0 & 0 & 0
		\end{pmatrix},
\end{equation}

\begin{equation}
\tilde{S}_{i\in C,D}^y= \mathcal{V}_{i\in C,D}^\dagger S_{i\in C,D}^y \mathcal{V}_{\in C,D}=\frac{1}{\sqrt{2}}
		\begin{pmatrix}
			0 & 0 & 1 \\
			0 & 0 & -i \\
			1 & i & 0
		\end{pmatrix},
\end{equation}

\begin{equation}
\tilde{S}_{i\in C,D}^z= \mathcal{V}_{i\in C,D}^\dagger S_{i\in C,D}^z \mathcal{V}_{\in C,D}=\frac{1}{\sqrt{2}}
		\begin{pmatrix}
			0 & 0 & i \\
			0 & 0 & -1 \\
			-i & -1 & 0
		\end{pmatrix}.
\end{equation}

In order to obtain the approximate structure factors up to quadratic order in the bosonic operators, we take the following expressions for each component of spin:

\begin{subequations}
	\begin{align}
		S^x_{A(B),\bm{q}} &\simeq 1, \\
		S^y_{A(B),\bm{q}} &\simeq -\frac{1}{\sqrt{2}} \left( \tilde{b}_{A(B),-\bm{q},2}^\dagger +\tilde{b}_{A(B),\bm{q},2} \right), \\
		S^z_{A(B),\bm{q}} &\simeq \frac{i}{\sqrt{2}}  \left( \tilde{b}_{A(B),-\bm{q},2}^\dagger -\tilde{b}_{A(B),\bm{q},2}  \right),
	\end{align}
\end{subequations}

\begin{subequations}
	\begin{align}
		S^x_{C(D),\bm{q}} &\simeq -1, \\
		S^y_{C(D),\bm{q}} &\simeq \frac{1}{\sqrt{2}} \left( \tilde{b}_{C(D),-\bm{q},2}^\dagger +\tilde{b}_{C(D),\bm{q},2} \right), \\
		S^z_{C(D),\bm{q}} &\simeq \frac{i}{\sqrt{2}}  \left( \tilde{b}_{C(D),-\bm{q},2}^\dagger -\tilde{b}_{C(D),\bm{q},2}  \right).
	\end{align}
\end{subequations}

Finally, the structure factors are given by the following expressions:

\begin{subequations}
	\begin{align}
		&\chi_{xx}^{\prime \prime} (\bm{q},\omega) = 1,\\
		&\chi_{yy}^{\prime \prime} (\bm{q},\omega) = \frac{1}{8}\sum_{i=1,4}|(V_{\vq}^{1i}+V_{\vq}^{2i})-(V_{\vq}^{3i}+V_{\vq}^{4i})|^2\delta(\omega-\omega_{\vq,2,+})\nonumber\\
		&+\frac{1}{8}\sum_{i=2,3}|(V_{\vq}^{1i}+V_{\vq}^{2i})-(V_{\vq}^{3i}+V_{\vq}^{4i})|^2\delta(\omega-\omega_{\vq,2,-}), \\
		&\chi_{zz}^{\prime \prime} (\bm{q},\omega) = \chi_{yy}^{\prime \prime} (\bm{q},\omega).
	\end{align}
\end{subequations}

\subsection{Flavor Wave for SD}

Just like before, all the symbolic expressions are the same as those in the section above, with the coefficients of the matrix in Eq.~(\ref{eq.matrix}) given by

\begin{subequations}
	\begin{align}
		\mathcal{J}_{11}&=\mathcal{J}_{22}=\mathcal{J}_{33}=\mathcal{J}_{44}=2J_1-K_1\equiv\mathcal{A}, \\
		\mathcal{J}_{12}&=\mathcal{J}_{14}^*=\mathcal{J}_{21}^*=\mathcal{J}_{23}=\mathcal{J}_{32}^*=\mathcal{J}_{34}=\mathcal{J}_{41}=\mathcal{J}_{43}^*\nonumber \\
		&=\frac{K_1}{2}e^{iq_x}+K_2e^{-iq_x}\cos{q_y}, \\
		\mathcal{J}_{13}&=\mathcal{J}_{24}=\mathcal{J}_{31}=\mathcal{J}_{42}=K_1\cos{(q_y)}\equiv\mathcal{B},
	\end{align}
\end{subequations}
and the coefficients $K_{ij}$ taking on the form
\begin{subequations}
	\begin{align}
		\mathcal{K}_{11}&=\mathcal{K}_{22}=\mathcal{K}_{33}=\mathcal{K}_{44}=J_1-\frac{3K_1}{2}-K_2\nonumber\\
		&+J_3\cos{(2q_y)}\equiv\mathcal{C}, \\
		\mathcal{K}_{12}&=\mathcal{K}_{21}^*=\mathcal{K}_{34}=\mathcal{K}_{43}^*=\frac{J_1}{2}e^{iq_x}+J_2e^{-iq_x}\cos{q_y}, \\
		\mathcal{K}_{14}^*&=\mathcal{K}_{23}=\mathcal{K}_{32}^*=\mathcal{K}_{41}\nonumber \\
		 &=-\frac{(J_1-K_1)}{2}e^{iq_x}-(J_2-K_2)e^{-iq_x}\cos{q_y}, \\
		\mathcal{K}_{13}&=\mathcal{K}_{24}=\mathcal{K}_{31}=\mathcal{K}_{42}\nonumber \\
		&=-(J_1-K_1)\cos{q_y}-J_3\cos{(2q_x)}\equiv\mathcal{D}.
	\end{align}
\end{subequations}

The constants contributing to the energy are now given by $E_0=-J_1+\frac{7}{2}K_1+3K_2$, and after diagonalizing, the resulting dispersions are now

\begin{subequations}
	\begin{align}
		\omega_{\vq,1,\pm} =&2\sqrt{\mathcal{A}^2-\mathcal{B}^2\pm\sqrt{\kappa_1}},\\
		\omega_{\vq,2,\pm} =&2\sqrt{\mathcal{C}^2-\mathcal{D}^2-\frac{K_1}{4}(K_1-2J_1)}\nonumber\\
		&\overline{-K_2(K_2-2J_2)\cos^2{q_y}\pm\sqrt{\kappa_2}},
	\end{align}
\end{subequations}

with

\begin{equation}
	\begin{aligned}
		\kappa_1=&(\mathcal{A}^2+\mathcal{B}^2)(K_1^2+4K_2^2\cos^2{q_y})\\
		&-\frac{1}{2}(K_1^2-4K_2^2\cos^2{q_y})^2[1-\cos{(4q_x)}]-8\mathcal{A}\mathcal{B}K_1K_2\cos{q_y}\\
		&+2(\mathcal{A}K_1-2\mathcal{B}K_2\cos{q_y})\\
		&(2\mathcal{A}K_2\cos{q_y}-\mathcal{B}K_1)\cos{(2q_x)},
	\end{aligned}
\end{equation}

\begin{equation}
	\begin{aligned}
		\kappa_2=&\mathcal{C}^2(J_1^2+4J_2^2\cos^2{q_y})+\mathcal{D}^2[(J_1-K_1)^2\\
		&+4(J_2-K_2)^2\cos^2{q_y}]\\
		&-\frac{1}{2}[J_1(J_1-K_1)-4J_2(J_2-K_2)\cos^2{q_y}]^2[1-\cos{(4q_x)}]\\
		&+4\mathcal{C}\mathcal{D}[J_1(J_2-K_2)+J_2(J_1-K_1)]\cos{q_y}\\
		&-2[\mathcal{C}J_1+2\mathcal{D}(J_2-K_2)\cos{q_y}]\\
		&[2\mathcal{C}J_2\cos{q_y}+\mathcal{D}(J_1-K_1)]\cos{(2q_x)},
	\end{aligned}
\end{equation}

where, as always, we write the resulting diagonalized Hamiltonian in the following form:

\begin{equation}
\begin{aligned}
\mathcal{H}_\text{fw}&=\sum_{\sigma=\pm}\sum_{a=1,2}\sum_{\bm{q}}\omega_{\bm{q},a,\sigma} (\alpha_{\bm{q},a,\sigma}^\dagger \alpha_{\bm{q},a,\sigma}\\ &+\beta_{\bm{q},a,\sigma}^\dagger \beta_{\bm{q},a,\sigma} + 1)\\
&+ N \left(E_0-3J_1+\frac{5}{2}K_1+K_2\right).
\end{aligned}
\end{equation}

\subsection{Flavor Wave for PL}

Unlike in the previous two cases, four sublattices are not enough to accurately describe the PL state and we must introduce four additional ones. The Hamiltonian matrix is still block diagonal with 8$\times$8 block matrices and where: $\psi^{\dagger}_{aa}=(b_{A,\bm{q},a},b_{B,\bm{q},a},b_{C,\bm{q},a},b_{D,\bm{q},a},b_{E,\bm{-q},a}^{\dagger},b_{F,\bm{-q},a}^{\dagger},b_{G,\bm{-q},a}^{\dagger},b_{H,\bm{-q},a}^{\dagger})$. The constants of the Hamiltonian are the same as on the case of the DS phase and the matrix elements are given by

\begin{subequations}
	\begin{align}
	\mathcal{K}_{11}&=\mathcal{K}_{22}=\mathcal{K}_{33}=\mathcal{K}_{44}\nonumber \\
	&=\mathcal{K}_{55}=\mathcal{K}_{66}=\mathcal{K}_{77}=\mathcal{K}_{88}=4J_3, \\
	\mathcal{K}_{15}&=\mathcal{K}_{26}=\mathcal{K}_{37}=\mathcal{K}_{48}\nonumber \\
	&=\mathcal{K}_{51}=\mathcal{K}_{62}=\mathcal{K}_{73}=\mathcal{K}_{84}=0, \\
	\mathcal{K}_{13}&=\mathcal{K}_{28}=\mathcal{K}_{31}=\mathcal{K}_{46}\nonumber \\
	&=\mathcal{K}_{57}=\mathcal{K}_{64}=\mathcal{K}_{75}=\mathcal{K}_{82}=K_2\cos{(q_x+q_y)}, \\
	\mathcal{K}_{17}&=\mathcal{K}_{24}=\mathcal{K}_{35}=\mathcal{K}_{42}\nonumber \\
	&=\mathcal{K}_{53}=\mathcal{K}_{68}=\mathcal{K}_{71}=\mathcal{K}_{86}=K_2\cos{(q_x-q_y)}, \\
	\mathcal{K}_{12}&=\mathcal{K}_{25}=\mathcal{K}_{38}=\mathcal{K}_{43}\nonumber \\
	&=\mathcal{K}_{56}=\mathcal{K}_{61}=\mathcal{K}_{74}=\mathcal{K}_{87}=\frac{K_1}{2}e^{iq_x}, \\
	\mathcal{K}_{16}&=\mathcal{K}_{21}=\mathcal{K}_{34}=\mathcal{K}_{47}\nonumber \\
	&=\mathcal{K}_{52}=\mathcal{K}_{65}=\mathcal{K}_{78}=\mathcal{K}_{83}=\frac{K_1}{2}e^{-iq_x}, \\
	\mathcal{K}_{14}&=\mathcal{K}_{23}=\mathcal{K}_{36}=\mathcal{K}_{45}\nonumber \\
	&=\mathcal{K}_{58}=\mathcal{K}_{67}=\mathcal{K}_{72}=\mathcal{K}_{81}=\frac{K_1}{2}e^{iq_y}, \\
	\mathcal{K}_{18}&=\mathcal{K}_{27}=\mathcal{K}_{32}=\mathcal{K}_{41}\nonumber \\
	&=\mathcal{K}_{54}=\mathcal{K}_{63}=\mathcal{K}_{76}=\mathcal{K}_{85}=\frac{K_1}{2}e^{-iq_y}, \\
	\end{align}
\end{subequations}

\begin{subequations}
	\begin{align}
	\mathcal{J}_{11}&=\mathcal{J}_{22}=\mathcal{J}_{33}=\mathcal{J}_{44}\nonumber \\
	&=\mathcal{J}_{55}=\mathcal{J}_{66}=\mathcal{J}_{77}=\mathcal{J}_{88}=2J_3-(K_1+K_2), \\
	\mathcal{J}_{15}&=\mathcal{J}_{26}=\mathcal{J}_{37}=\mathcal{J}_{48}=\mathcal{J}_{51}=\mathcal{J}_{62}=\mathcal{J}_{73}=\mathcal{J}_{84}\nonumber \\
	&=-J_3[\cos{(2q_x)}+\cos{(2q_y)}], \\
	\mathcal{J}_{13}&=\mathcal{J}_{31}=\mathcal{J}_{57}=\mathcal{J}_{75}=J_2\cos{(q_x+q_y)}, \\
	\mathcal{J}_{24}&=\mathcal{J}_{42}=\mathcal{J}_{68}=\mathcal{J}_{86}=J_2\cos{(q_x-q_y)}, \\
	\mathcal{J}_{28}&=\mathcal{J}_{46}=\mathcal{J}_{64}=\mathcal{J}_{82}=-(J_2-K_2)\cos{(q_x+q_y)}, \\
	\mathcal{J}_{17}&=\mathcal{J}_{35}=\mathcal{J}_{53}=\mathcal{J}_{71}=-(J_2-K_2)\cos{(q_x-q_y)}, \\
	\mathcal{J}_{12}&=\mathcal{J}_{43}=\mathcal{J}_{56}=\mathcal{J}_{87}=\frac{J_1}{2}e^{iq_x}, \\
	\mathcal{J}_{21}&=\mathcal{J}_{34}=\mathcal{J}_{65}=\mathcal{J}_{78}=\frac{J_1}{2}e^{-iq_x}, \\
	\mathcal{J}_{14}&=\mathcal{J}_{23}=\mathcal{J}_{58}=\mathcal{J}_{67}=\frac{J_1}{2}e^{iq_y}, \\
	\mathcal{J}_{32}&=\mathcal{J}_{41}=\mathcal{J}_{76}=\mathcal{J}_{85}=\frac{J_1}{2}e^{-iq_y}, \\
	\mathcal{J}_{25}&=\mathcal{J}_{38}=\mathcal{J}_{61}=\mathcal{J}_{74}=-\frac{(J_1-K_1)}{2}e^{iq_x}, \\
	\mathcal{J}_{16}&=\mathcal{J}_{47}=\mathcal{J}_{52}=\mathcal{J}_{83}=-\frac{(J_1-K_1)}{2}e^{-iq_x}, \\
	\mathcal{J}_{36}&=\mathcal{J}_{45}=\mathcal{J}_{72}=\mathcal{J}_{81}=-\frac{(J_1-K_1)}{2}e^{iq_y}, \\
	\mathcal{J}_{18}&=\mathcal{J}_{27}=\mathcal{J}_{54}=\mathcal{J}_{63}=-\frac{(J_1-K_1)}{2}e^{-iq_y}. \\
	\end{align}
\end{subequations}

Because of their complexity in this case, analytical expressions for the dispersions are not included in this case. However, these can be obtained using the technique described in Sec. V above. The Hamiltonian takes the following form:

\begin{equation}
\begin{aligned}
\mathcal{H}_\text{fw}&=\sum_{\sigma}\sum_{a=1,2}\sum_{\bm{q}}\omega_{\bm{q},a,\sigma} (\alpha_{\bm{q},a,\sigma}^\dagger \alpha_{\bm{q},a,\sigma} + \beta_{\bm{q},a,\sigma}^\dagger \beta_{\bm{q},a,\sigma} + 1)\\
&+ N (E_0+K_1+K_2-6J_3),
\end{aligned}
\end{equation}

where the index $\sigma$ is added in order to account for the summation over all the different dispersions obtained for each of the two modes.


\begin{thebibliography}{64}%
	\makeatletter
	\providecommand \@ifxundefined [1]{%
		\@ifx{#1\undefined}
	}%
	\providecommand \@ifnum [1]{%
		\ifnum #1\expandafter \@firstoftwo
		\else \expandafter \@secondoftwo
		\fi
	}%
	\providecommand \@ifx [1]{%
		\ifx #1\expandafter \@firstoftwo
		\else \expandafter \@secondoftwo
		\fi
	}%
	\providecommand \natexlab [1]{#1}%
	\providecommand \enquote  [1]{``#1''}%
	\providecommand \bibnamefont  [1]{#1}%
	\providecommand \bibfnamefont [1]{#1}%
	\providecommand \citenamefont [1]{#1}%
	\providecommand \href@noop [0]{\@secondoftwo}%
	\providecommand \href [0]{\begingroup \@sanitize@url \@href}%
	\providecommand \@href[1]{\@@startlink{#1}\@@href}%
	\providecommand \@@href[1]{\endgroup#1\@@endlink}%
	\providecommand \@sanitize@url [0]{\catcode `\\12\catcode `\$12\catcode
		`\&12\catcode `\#12\catcode `\^12\catcode `\_12\catcode `\%12\relax}%
	\providecommand \@@startlink[1]{}%
	\providecommand \@@endlink[0]{}%
	\providecommand \url  [0]{\begingroup\@sanitize@url \@url }%
	\providecommand \@url [1]{\endgroup\@href {#1}{\urlprefix }}%
	\providecommand \urlprefix  [0]{URL }%
	\providecommand \Eprint [0]{\href }%
	\providecommand \doibase [0]{http://dx.doi.org/}%
	\providecommand \selectlanguage [0]{\@gobble}%
	\providecommand \bibinfo  [0]{\@secondoftwo}%
	\providecommand \bibfield  [0]{\@secondoftwo}%
	\providecommand \translation [1]{[#1]}%
	\providecommand \BibitemOpen [0]{}%
	\providecommand \bibitemStop [0]{}%
	\providecommand \bibitemNoStop [0]{.\EOS\space}%
	\providecommand \EOS [0]{\spacefactor3000\relax}%
	\providecommand \BibitemShut  [1]{\csname bibitem#1\endcsname}%
	\let\auto@bib@innerbib\@empty
	\bibitem [{\citenamefont {Bao}\ \emph {et~al.}(2009)\citenamefont {Bao},
		\citenamefont {Qiu}, \citenamefont {Huang}, \citenamefont {Green},
		\citenamefont {Zajdel}, \citenamefont {Fitzsimmons}, \citenamefont
		{Zhernenkov}, \citenamefont {Chang}, \citenamefont {Fang}, \citenamefont
		{Qian}, \citenamefont {Vehstedt}, \citenamefont {Yang}, \citenamefont {Pham},
		\citenamefont {Spinu},\ and\ \citenamefont {Mao}}]{Bao2009}%
	\BibitemOpen
	\bibfield  {author} {\bibinfo {author} {\bibfnamefont {W.}~\bibnamefont
			{Bao}}, \bibinfo {author} {\bibfnamefont {Y.}~\bibnamefont {Qiu}}, \bibinfo
		{author} {\bibfnamefont {Q.}~\bibnamefont {Huang}}, \bibinfo {author}
		{\bibfnamefont {M.~A.}\ \bibnamefont {Green}}, \bibinfo {author}
		{\bibfnamefont {P.}~\bibnamefont {Zajdel}}, \bibinfo {author} {\bibfnamefont
			{M.~R.}\ \bibnamefont {Fitzsimmons}}, \bibinfo {author} {\bibfnamefont
			{M.}~\bibnamefont {Zhernenkov}}, \bibinfo {author} {\bibfnamefont
			{S.}~\bibnamefont {Chang}}, \bibinfo {author} {\bibfnamefont
			{M.}~\bibnamefont {Fang}}, \bibinfo {author} {\bibfnamefont {B.}~\bibnamefont
			{Qian}}, \bibinfo {author} {\bibfnamefont {E.~K.}\ \bibnamefont {Vehstedt}},
		\bibinfo {author} {\bibfnamefont {J.}~\bibnamefont {Yang}}, \bibinfo {author}
		{\bibfnamefont {H.~M.}\ \bibnamefont {Pham}}, \bibinfo {author}
		{\bibfnamefont {L.}~\bibnamefont {Spinu}}, \ and\ \bibinfo {author}
		{\bibfnamefont {Z.~Q.}\ \bibnamefont {Mao}},\ }\href@noop {} {\bibfield
		{journal} {\bibinfo  {journal} {Phys. Rev. Lett.}\ }\textbf {\bibinfo
			{volume} {102}},\ \bibinfo {pages} {247001} (\bibinfo {year}
		{2009})}\BibitemShut {NoStop}%
	\bibitem [{\citenamefont {Li}\ \emph {et~al.}(2009)\citenamefont {Li},
		\citenamefont {delaCruz}, \citenamefont {Huang}, \citenamefont {Chen},
		\citenamefont {Lynn}, \citenamefont {Hu}, \citenamefont {Huang},
		\citenamefont {Hsu}, \citenamefont {Yeh}, \citenamefont {Wu},\ and\
		\citenamefont {Dai}}]{Li2009}%
	\BibitemOpen
	\bibfield  {author} {\bibinfo {author} {\bibfnamefont {S.}~\bibnamefont
			{Li}}, \bibinfo {author} {\bibfnamefont {C.}~\bibnamefont {delaCruz}},
		\bibinfo {author} {\bibfnamefont {Q.}~\bibnamefont {Huang}}, \bibinfo
		{author} {\bibfnamefont {Y.}~\bibnamefont {Chen}}, \bibinfo {author}
		{\bibfnamefont {J.~W.}\ \bibnamefont {Lynn}}, \bibinfo {author}
		{\bibfnamefont {J.}~\bibnamefont {Hu}}, \bibinfo {author} {\bibfnamefont
			{Y.~L.}\ \bibnamefont {Huang}}, \bibinfo {author} {\bibfnamefont {F.~C.}\
			\bibnamefont {Hsu}}, \bibinfo {author} {\bibfnamefont {K.~W.}\ \bibnamefont
			{Yeh}}, \bibinfo {author} {\bibfnamefont {M.~K.}\ \bibnamefont {Wu}}, \ and\
		\bibinfo {author} {\bibfnamefont {P.}~\bibnamefont {Dai}},\ }\href@noop {}
	{\bibfield  {journal} {\bibinfo  {journal} {Phys. Rev. B}\ }\textbf {\bibinfo
			{volume} {79}},\ \bibinfo {pages} {054503} (\bibinfo {year}
		{2009})}\BibitemShut {NoStop}%
	\bibitem [{\citenamefont {Wen}\ \emph {et~al.}(2009)\citenamefont {Wen},
		\citenamefont {Xu}, \citenamefont {Xu}, \citenamefont {Lin}, \citenamefont
		{Li}, \citenamefont {Ratcliff}, \citenamefont {Gu},\ and\ \citenamefont
		{Tranquada}}]{Wen2009}%
	\BibitemOpen
	\bibfield  {author} {\bibinfo {author} {\bibfnamefont {J.}~\bibnamefont
			{Wen}}, \bibinfo {author} {\bibfnamefont {G.}~\bibnamefont {Xu}}, \bibinfo
		{author} {\bibfnamefont {Z.}~\bibnamefont {Xu}}, \bibinfo {author}
		{\bibfnamefont {Z.~W.}\ \bibnamefont {Lin}}, \bibinfo {author} {\bibfnamefont
			{Q.}~\bibnamefont {Li}}, \bibinfo {author} {\bibfnamefont {W.}~\bibnamefont
			{Ratcliff}}, \bibinfo {author} {\bibfnamefont {G.}~\bibnamefont {Gu}}, \ and\
		\bibinfo {author} {\bibfnamefont {J.~M.}\ \bibnamefont {Tranquada}},\
	}\href@noop {} {\bibfield  {journal} {\bibinfo  {journal} {Phys. Rev. B}\
		}\textbf {\bibinfo {volume} {80}},\ \bibinfo {pages} {104506} (\bibinfo
		{year} {2009})}\BibitemShut {NoStop}%
	\bibitem [{\citenamefont {de~la Cruz}\ \emph {et~al.}(2008)\citenamefont {de~la
			Cruz}, \citenamefont {Huang}, \citenamefont {Lynn}, \citenamefont {Li},
		\citenamefont {II}, \citenamefont {Zarestky}, \citenamefont {Mook},
		\citenamefont {Chen}, \citenamefont {Luo}, \citenamefont {Wang},\ and\
		\citenamefont {Dai}}]{LaCruz2008}%
	\BibitemOpen
	\bibfield  {author} {\bibinfo {author} {\bibfnamefont {C.}~\bibnamefont
			{de~la Cruz}}, \bibinfo {author} {\bibfnamefont {Q.}~\bibnamefont {Huang}},
		\bibinfo {author} {\bibfnamefont {J.~W.}\ \bibnamefont {Lynn}}, \bibinfo
		{author} {\bibfnamefont {J.}~\bibnamefont {Li}}, \bibinfo {author}
		{\bibfnamefont {W.~R.}\ \bibnamefont {II}}, \bibinfo {author} {\bibfnamefont
			{J.~L.}\ \bibnamefont {Zarestky}}, \bibinfo {author} {\bibfnamefont {H.~A.}\
			\bibnamefont {Mook}}, \bibinfo {author} {\bibfnamefont {G.~F.}\ \bibnamefont
			{Chen}}, \bibinfo {author} {\bibfnamefont {J.~L.}\ \bibnamefont {Luo}},
		\bibinfo {author} {\bibfnamefont {N.~L.}\ \bibnamefont {Wang}}, \ and\
		\bibinfo {author} {\bibfnamefont {P.}~\bibnamefont {Dai}},\ }\href
	{http://dx.doi.org/10.1038/nature07057} {\bibfield  {journal} {\bibinfo
			{journal} {Nature}\ }\textbf {\bibinfo {volume} {453}},\ \bibinfo {pages}
		{899} (\bibinfo {year} {2008})}\BibitemShut {NoStop}%
	\bibitem [{\citenamefont {Lumsden}\ and\ \citenamefont
		{Christianson}(2010)}]{Lumsden2010}%
	\BibitemOpen
	\bibfield  {author} {\bibinfo {author} {\bibfnamefont {M.~D.}\ \bibnamefont
			{Lumsden}}\ and\ \bibinfo {author} {\bibfnamefont {A.~D.}\ \bibnamefont
			{Christianson}},\ }\href {http://stacks.iop.org/0953-8984/22/i=20/a=203203}
	{\bibfield  {journal} {\bibinfo  {journal} {J. Phys.: Condensed Matter}\
		}\textbf {\bibinfo {volume} {22}},\ \bibinfo {pages} {203203} (\bibinfo
		{year} {2010})}\BibitemShut {NoStop}%
	\bibitem [{\citenamefont {Dai}(2015)}]{Dai2015}%
	\BibitemOpen
	\bibfield  {author} {\bibinfo {author} {\bibfnamefont {P.}~\bibnamefont
			{Dai}},\ }\href {\doibase 10.1103/RevModPhys.87.855} {\bibfield  {journal}
		{\bibinfo  {journal} {Rev. Mod. Phys.}\ }\textbf {\bibinfo {volume} {87}},\
		\bibinfo {pages} {855} (\bibinfo {year} {2015})}\BibitemShut {NoStop}%
	\bibitem [{\citenamefont {Qiu}\ \emph {et~al.}(2009)\citenamefont {Qiu},
		\citenamefont {Bao}, \citenamefont {Zhao}, \citenamefont {Broholm},
		\citenamefont {Stanev}, \citenamefont {Tesanovic}, \citenamefont
		{Gasparovic}, \citenamefont {Chang}, \citenamefont {Hu}, \citenamefont
		{Qian}, \citenamefont {Fang},\ and\ \citenamefont {Mao}}]{Qiu2009}%
	\BibitemOpen
	\bibfield  {author} {\bibinfo {author} {\bibfnamefont {Y.}~\bibnamefont
			{Qiu}}, \bibinfo {author} {\bibfnamefont {W.}~\bibnamefont {Bao}}, \bibinfo
		{author} {\bibfnamefont {Y.}~\bibnamefont {Zhao}}, \bibinfo {author}
		{\bibfnamefont {C.}~\bibnamefont {Broholm}}, \bibinfo {author} {\bibfnamefont
			{V.}~\bibnamefont {Stanev}}, \bibinfo {author} {\bibfnamefont
			{Z.}~\bibnamefont {Tesanovic}}, \bibinfo {author} {\bibfnamefont {Y.~C.}\
			\bibnamefont {Gasparovic}}, \bibinfo {author} {\bibfnamefont
			{S.}~\bibnamefont {Chang}}, \bibinfo {author} {\bibfnamefont
			{J.}~\bibnamefont {Hu}}, \bibinfo {author} {\bibfnamefont {B.}~\bibnamefont
			{Qian}}, \bibinfo {author} {\bibfnamefont {M.}~\bibnamefont {Fang}}, \ and\
		\bibinfo {author} {\bibfnamefont {Z.}~\bibnamefont {Mao}},\ }\href@noop {}
	{\bibfield  {journal} {\bibinfo  {journal} {Phys. Rev. Lett.}\ }\textbf
		{\bibinfo {volume} {103}},\ \bibinfo {pages} {067008} (\bibinfo {year}
		{2009})}\BibitemShut {NoStop}%
	\bibitem [{\citenamefont {Lee}\ \emph {et~al.}(2010)\citenamefont {Lee},
		\citenamefont {Xu}, \citenamefont {Ku}, \citenamefont {Wen}, \citenamefont
		{Lee}, \citenamefont {Katayama}, \citenamefont {Xu}, \citenamefont {Ji},
		\citenamefont {Lin}, \citenamefont {Gu}, \citenamefont {Yang}, \citenamefont
		{Johnson}, \citenamefont {Pan}, \citenamefont {Valla}, \citenamefont
		{Fujita}, \citenamefont {Sato}, \citenamefont {Chang}, \citenamefont
		{Yamada},\ and\ \citenamefont {Tranquada}}]{Lee2010}%
	\BibitemOpen
	\bibfield  {author} {\bibinfo {author} {\bibfnamefont {S.~H.}\ \bibnamefont
			{Lee}}, \bibinfo {author} {\bibfnamefont {G.}~\bibnamefont {Xu}}, \bibinfo
		{author} {\bibfnamefont {W.}~\bibnamefont {Ku}}, \bibinfo {author}
		{\bibfnamefont {J.~S.}\ \bibnamefont {Wen}}, \bibinfo {author} {\bibfnamefont
			{C.~C.}\ \bibnamefont {Lee}}, \bibinfo {author} {\bibfnamefont
			{N.}~\bibnamefont {Katayama}}, \bibinfo {author} {\bibfnamefont {Z.~J.}\
			\bibnamefont {Xu}}, \bibinfo {author} {\bibfnamefont {S.}~\bibnamefont {Ji}},
		\bibinfo {author} {\bibfnamefont {Z.~W.}\ \bibnamefont {Lin}}, \bibinfo
		{author} {\bibfnamefont {G.~D.}\ \bibnamefont {Gu}}, \bibinfo {author}
		{\bibfnamefont {H.~B.}\ \bibnamefont {Yang}}, \bibinfo {author}
		{\bibfnamefont {P.~D.}\ \bibnamefont {Johnson}}, \bibinfo {author}
		{\bibfnamefont {Z.~H.}\ \bibnamefont {Pan}}, \bibinfo {author} {\bibfnamefont
			{T.}~\bibnamefont {Valla}}, \bibinfo {author} {\bibfnamefont
			{M.}~\bibnamefont {Fujita}}, \bibinfo {author} {\bibfnamefont {T.~J.}\
			\bibnamefont {Sato}}, \bibinfo {author} {\bibfnamefont {S.}~\bibnamefont
			{Chang}}, \bibinfo {author} {\bibfnamefont {K.}~\bibnamefont {Yamada}}, \
		and\ \bibinfo {author} {\bibfnamefont {J.~M.}\ \bibnamefont {Tranquada}},\
	}\href@noop {} {\bibfield  {journal} {\bibinfo  {journal} {Phys. Rev. B}\
		}\textbf {\bibinfo {volume} {81}},\ \bibinfo {pages} {220502} (\bibinfo
		{year} {2010})}\BibitemShut {NoStop}%
	\bibitem [{\citenamefont {Lumsden~{\it et al.}}(2010)}]{Lumsden_NPhys2010}%
	\BibitemOpen
	\bibfield  {author} {\bibinfo {author} {\bibfnamefont {M.~D.}\ \bibnamefont
			{Lumsden~{\it et al.}}},\ }\href@noop {} {\bibfield  {journal} {\bibinfo
			{journal} {Nat. Phys.}\ }\textbf {\bibinfo {volume} {6}},\ \bibinfo {pages}
		{182} (\bibinfo {year} {2010})}\BibitemShut {NoStop}%
	\bibitem [{\citenamefont {Liu~{\it et al.}}(2010)}]{Liu2010}%
	\BibitemOpen
	\bibfield  {author} {\bibinfo {author} {\bibfnamefont {T.~J.}\ \bibnamefont
			{Liu~{\it et al.}}},\ }\href@noop {} {\bibfield  {journal} {\bibinfo
			{journal} {Nat. Mater.}\ }\textbf {\bibinfo {volume} {9}},\ \bibinfo {pages}
		{718} (\bibinfo {year} {2010})}\BibitemShut {NoStop}%
	\bibitem [{\citenamefont {Xu}\ \emph {et~al.}(2016)\citenamefont {Xu},
		\citenamefont {Schneeloch}, \citenamefont {Wen}, \citenamefont {Bozin},
		\citenamefont {Granroth}, \citenamefont {Winn}, \citenamefont {Feygenson},
		\citenamefont {Birgeneau}, \citenamefont {Gu}, \citenamefont {Zaliznyak},
		\citenamefont {Tranquada},\ and\ \citenamefont {Xu}}]{Xu2016}%
	\BibitemOpen
	\bibfield  {author} {\bibinfo {author} {\bibfnamefont {Z.}~\bibnamefont
			{Xu}}, \bibinfo {author} {\bibfnamefont {J.~A.}\ \bibnamefont {Schneeloch}},
		\bibinfo {author} {\bibfnamefont {J.}~\bibnamefont {Wen}}, \bibinfo {author}
		{\bibfnamefont {E.~S.}\ \bibnamefont {Bozin}}, \bibinfo {author}
		{\bibfnamefont {G.~E.}\ \bibnamefont {Granroth}}, \bibinfo {author}
		{\bibfnamefont {B.~L.}\ \bibnamefont {Winn}}, \bibinfo {author}
		{\bibfnamefont {M.}~\bibnamefont {Feygenson}}, \bibinfo {author}
		{\bibfnamefont {R.~J.}\ \bibnamefont {Birgeneau}}, \bibinfo {author}
		{\bibfnamefont {G.}~\bibnamefont {Gu}}, \bibinfo {author} {\bibfnamefont
			{I.~A.}\ \bibnamefont {Zaliznyak}}, \bibinfo {author} {\bibfnamefont {J.~M.}\
			\bibnamefont {Tranquada}}, \ and\ \bibinfo {author} {\bibfnamefont
			{G.}~\bibnamefont {Xu}},\ }\href@noop {} {\bibfield  {journal} {\bibinfo
			{journal} {Phys. Rev. B}\ }\textbf {\bibinfo {volume} {93}},\ \bibinfo
		{pages} {104517} (\bibinfo {year} {2016})}\BibitemShut {NoStop}%
	\bibitem [{\citenamefont {McQueen}\ \emph
		{et~al.}(2009{\natexlab{a}})\citenamefont {McQueen}, \citenamefont {Huang},
		\citenamefont {Ksenofontov}, \citenamefont {Felser}, \citenamefont {Xu},
		\citenamefont {Zandbergen}, \citenamefont {Hor}, \citenamefont {Allred},
		\citenamefont {Williams}, \citenamefont {Qu}, \citenamefont {Checkelsky},
		\citenamefont {Ong},\ and\ \citenamefont {Cava}}]{McQueen2009}%
	\BibitemOpen
	\bibfield  {author} {\bibinfo {author} {\bibfnamefont {T.~M.}\ \bibnamefont
			{McQueen}}, \bibinfo {author} {\bibfnamefont {Q.}~\bibnamefont {Huang}},
		\bibinfo {author} {\bibfnamefont {V.}~\bibnamefont {Ksenofontov}}, \bibinfo
		{author} {\bibfnamefont {C.}~\bibnamefont {Felser}}, \bibinfo {author}
		{\bibfnamefont {Q.}~\bibnamefont {Xu}}, \bibinfo {author} {\bibfnamefont
			{H.}~\bibnamefont {Zandbergen}}, \bibinfo {author} {\bibfnamefont {Y.~S.}\
			\bibnamefont {Hor}}, \bibinfo {author} {\bibfnamefont {J.}~\bibnamefont
			{Allred}}, \bibinfo {author} {\bibfnamefont {A.~J.}\ \bibnamefont
			{Williams}}, \bibinfo {author} {\bibfnamefont {D.}~\bibnamefont {Qu}},
		\bibinfo {author} {\bibfnamefont {J.}~\bibnamefont {Checkelsky}}, \bibinfo
		{author} {\bibfnamefont {N.~P.}\ \bibnamefont {Ong}}, \ and\ \bibinfo
		{author} {\bibfnamefont {R.~J.}\ \bibnamefont {Cava}},\ }\href {\doibase
		10.1103/PhysRevB.79.014522} {\bibfield  {journal} {\bibinfo  {journal} {Phys.
				Rev. B}\ }\textbf {\bibinfo {volume} {79}},\ \bibinfo {pages} {014522}
		(\bibinfo {year} {2009}{\natexlab{a}})}\BibitemShut {NoStop}%
	\bibitem [{\citenamefont {Bendele}\ \emph {et~al.}(2010)\citenamefont
		{Bendele}, \citenamefont {Amato}, \citenamefont {Conder}, \citenamefont
		{Elender}, \citenamefont {Keller}, \citenamefont {Klauss}, \citenamefont
		{Luetkens}, \citenamefont {Pomjakushina}, \citenamefont {Raselli},\ and\
		\citenamefont {Khasanov}}]{Bendele2010}%
	\BibitemOpen
	\bibfield  {author} {\bibinfo {author} {\bibfnamefont {M.}~\bibnamefont
			{Bendele}}, \bibinfo {author} {\bibfnamefont {A.}~\bibnamefont {Amato}},
		\bibinfo {author} {\bibfnamefont {K.}~\bibnamefont {Conder}}, \bibinfo
		{author} {\bibfnamefont {M.}~\bibnamefont {Elender}}, \bibinfo {author}
		{\bibfnamefont {H.}~\bibnamefont {Keller}}, \bibinfo {author} {\bibfnamefont
			{H.-H.}\ \bibnamefont {Klauss}}, \bibinfo {author} {\bibfnamefont
			{H.}~\bibnamefont {Luetkens}}, \bibinfo {author} {\bibfnamefont
			{E.}~\bibnamefont {Pomjakushina}}, \bibinfo {author} {\bibfnamefont
			{A.}~\bibnamefont {Raselli}}, \ and\ \bibinfo {author} {\bibfnamefont
			{R.}~\bibnamefont {Khasanov}},\ }\href {\doibase
		10.1103/PhysRevLett.104.087003} {\bibfield  {journal} {\bibinfo  {journal}
			{Phys. Rev. Lett.}\ }\textbf {\bibinfo {volume} {104}},\ \bibinfo {pages}
		{087003} (\bibinfo {year} {2010})}\BibitemShut {NoStop}%
	\bibitem [{\citenamefont {Rahn}\ \emph {et~al.}(2015)\citenamefont {Rahn},
		\citenamefont {Ewings}, \citenamefont {Sedlmaier}, \citenamefont {Clarke},\
		and\ \citenamefont {Boothroyd}}]{Rahn2015}%
	\BibitemOpen
	\bibfield  {author} {\bibinfo {author} {\bibfnamefont {M.~C.}\ \bibnamefont
			{Rahn}}, \bibinfo {author} {\bibfnamefont {R.~A.}\ \bibnamefont {Ewings}},
		\bibinfo {author} {\bibfnamefont {S.~J.}\ \bibnamefont {Sedlmaier}}, \bibinfo
		{author} {\bibfnamefont {S.~J.}\ \bibnamefont {Clarke}}, \ and\ \bibinfo
		{author} {\bibfnamefont {A.~T.}\ \bibnamefont {Boothroyd}},\ }\href {\doibase
		10.1103/PhysRevB.91.180501} {\bibfield  {journal} {\bibinfo  {journal} {Phys.
				Rev. B}\ }\textbf {\bibinfo {volume} {91}},\ \bibinfo {pages} {180501}
		(\bibinfo {year} {2015})}\BibitemShut {NoStop}%
	\bibitem [{\citenamefont {Wang}\ \emph
		{et~al.}(2015{\natexlab{a}})\citenamefont {Wang}, \citenamefont {Shen},
		\citenamefont {Pan}, \citenamefont {Zhang}, \citenamefont {Ikeuchi},
		\citenamefont {Iida}, \citenamefont {Christianson}, \citenamefont {Walker},
		\citenamefont {Adroja}, \citenamefont {Abdel-Hafiez}, \citenamefont {Chen},
		\citenamefont {Chareev}, \citenamefont {Vasiliev},\ and\ \citenamefont
		{Zhao}}]{QWang2015}%
	\BibitemOpen
	\bibfield  {author} {\bibinfo {author} {\bibfnamefont {Q.}~\bibnamefont
			{Wang}}, \bibinfo {author} {\bibfnamefont {Y.}~\bibnamefont {Shen}}, \bibinfo
		{author} {\bibfnamefont {B.}~\bibnamefont {Pan}}, \bibinfo {author}
		{\bibfnamefont {X.}~\bibnamefont {Zhang}}, \bibinfo {author} {\bibfnamefont
			{K.}~\bibnamefont {Ikeuchi}}, \bibinfo {author} {\bibfnamefont
			{K.}~\bibnamefont {Iida}}, \bibinfo {author} {\bibfnamefont {A.~D.}\
			\bibnamefont {Christianson}}, \bibinfo {author} {\bibfnamefont {H.~C.}\
			\bibnamefont {Walker}}, \bibinfo {author} {\bibfnamefont {D.~T.}\
			\bibnamefont {Adroja}}, \bibinfo {author} {\bibfnamefont {M.}~\bibnamefont
			{Abdel-Hafiez}}, \bibinfo {author} {\bibfnamefont {X.}~\bibnamefont {Chen}},
		\bibinfo {author} {\bibfnamefont {D.~A.}\ \bibnamefont {Chareev}}, \bibinfo
		{author} {\bibfnamefont {A.~N.}\ \bibnamefont {Vasiliev}}, \ and\ \bibinfo
		{author} {\bibfnamefont {J.}~\bibnamefont {Zhao}},\ }\href@noop {} {\bibfield  {journal} {\bibinfo  {journal} {Nat. Commun.}\ }\textbf {\bibinfo {volume} {7}},\ \bibinfo {pages} {12182}
		(\bibinfo {year} {2016})}\BibitemShut {NoStop}%
	\bibitem [{\citenamefont {Wang}\ \emph {et~al.}(2016)\citenamefont {Wang},
		\citenamefont {Shen}, \citenamefont {Pan}, \citenamefont {Hao}, \citenamefont
		{Ma}, \citenamefont {Zhou}, \citenamefont {Steffens}, \citenamefont
		{Schmalzl}, \citenamefont {Forrest}, \citenamefont {Abdel-Hafiez},
		\citenamefont {Chen}, \citenamefont {Chareev}, \citenamefont {Vasiliev},
		\citenamefont {Bourges}, \citenamefont {Sidis}, \citenamefont {Cao},\ and\
		\citenamefont {Zhao}}]{QWang2016}%
	\BibitemOpen
	\bibfield  {author} {\bibinfo {author} {\bibfnamefont {Q.}~\bibnamefont
			{Wang}}, \bibinfo {author} {\bibfnamefont {Y.}~\bibnamefont {Shen}}, \bibinfo
		{author} {\bibfnamefont {B.}~\bibnamefont {Pan}}, \bibinfo {author}
		{\bibfnamefont {Y.}~\bibnamefont {Hao}}, \bibinfo {author} {\bibfnamefont
			{M.}~\bibnamefont {Ma}}, \bibinfo {author} {\bibfnamefont {F.}~\bibnamefont
			{Zhou}}, \bibinfo {author} {\bibfnamefont {P.}~\bibnamefont {Steffens}},
		\bibinfo {author} {\bibfnamefont {K.}~\bibnamefont {Schmalzl}}, \bibinfo
		{author} {\bibfnamefont {T.~R.}\ \bibnamefont {Forrest}}, \bibinfo {author}
		{\bibfnamefont {M.}~\bibnamefont {Abdel-Hafiez}}, \bibinfo {author}
		{\bibfnamefont {X.}~\bibnamefont {Chen}}, \bibinfo {author} {\bibfnamefont
			{D.~A.}\ \bibnamefont {Chareev}}, \bibinfo {author} {\bibfnamefont {A.~N.}\
			\bibnamefont {Vasiliev}}, \bibinfo {author} {\bibfnamefont {P.}~\bibnamefont
			{Bourges}}, \bibinfo {author} {\bibfnamefont {Y.}~\bibnamefont {Sidis}},
		\bibinfo {author} {\bibfnamefont {H.}~\bibnamefont {Cao}}, \ and\ \bibinfo
		{author} {\bibfnamefont {J.}~\bibnamefont {Zhao}},\ }\href@noop {} {\bibfield
		{journal} {\bibinfo  {journal} {Nat Mater}\ }\textbf {\bibinfo {volume}
			{15}},\ \bibinfo {pages} {159} (\bibinfo {year} {2016})}\BibitemShut
	{NoStop}%
	\bibitem [{\citenamefont {Shamoto}\ \emph {et~al.}(2015)\citenamefont
		{Shamoto}, \citenamefont {Matsuoka}, \citenamefont {Kajimoto}, \citenamefont
		{Ishikado}, \citenamefont {Yamakawa}, \citenamefont {Watashige},
		\citenamefont {Kasahara}, \citenamefont {Nakamura}, \citenamefont {Kontani},
		\citenamefont {Shibauchi},\ and\ \citenamefont {Matsuda}}]{Shamoto2015}%
	\BibitemOpen
	\bibfield  {author} {\bibinfo {author} {\bibfnamefont {S.}~\bibnamefont
			{Shamoto}}, \bibinfo {author} {\bibfnamefont {K.}~\bibnamefont {Matsuoka}},
		\bibinfo {author} {\bibfnamefont {R.}~\bibnamefont {Kajimoto}}, \bibinfo
		{author} {\bibfnamefont {M.}~\bibnamefont {Ishikado}}, \bibinfo {author}
		{\bibfnamefont {Y.}~\bibnamefont {Yamakawa}}, \bibinfo {author}
		{\bibfnamefont {T.}~\bibnamefont {Watashige}}, \bibinfo {author}
		{\bibfnamefont {S.}~\bibnamefont {Kasahara}}, \bibinfo {author}
		{\bibfnamefont {M.}~\bibnamefont {Nakamura}}, \bibinfo {author}
		{\bibfnamefont {H.}~\bibnamefont {Kontani}}, \bibinfo {author} {\bibfnamefont
			{T.}~\bibnamefont {Shibauchi}}, \ and\ \bibinfo {author} {\bibfnamefont
			{Y.}~\bibnamefont {Matsuda}},\ }\href@noop {} {\bibfield  {journal} {\bibinfo
			{journal} {arXiv:1511.04267}\ } (\bibinfo {year} {2015})}\BibitemShut
	{NoStop}%
	\bibitem [{\citenamefont {Bendele}\ \emph {et~al.}(2012)\citenamefont
		{Bendele}, \citenamefont {Ichsanow}, \citenamefont {Pashkevich},
		\citenamefont {Keller}, \citenamefont {Str\"assle}, \citenamefont {Gusev},
		\citenamefont {Pomjakushina}, \citenamefont {Conder}, \citenamefont
		{Khasanov},\ and\ \citenamefont {Keller}}]{Bendele2012}%
	\BibitemOpen
	\bibfield  {author} {\bibinfo {author} {\bibfnamefont {M.}~\bibnamefont
			{Bendele}}, \bibinfo {author} {\bibfnamefont {A.}~\bibnamefont {Ichsanow}},
		\bibinfo {author} {\bibfnamefont {Y.}~\bibnamefont {Pashkevich}}, \bibinfo
		{author} {\bibfnamefont {L.}~\bibnamefont {Keller}}, \bibinfo {author}
		{\bibfnamefont {T.}~\bibnamefont {Str\"assle}}, \bibinfo {author}
		{\bibfnamefont {A.}~\bibnamefont {Gusev}}, \bibinfo {author} {\bibfnamefont
			{E.}~\bibnamefont {Pomjakushina}}, \bibinfo {author} {\bibfnamefont
			{K.}~\bibnamefont {Conder}}, \bibinfo {author} {\bibfnamefont
			{R.}~\bibnamefont {Khasanov}}, \ and\ \bibinfo {author} {\bibfnamefont
			{H.}~\bibnamefont {Keller}},\ }\href {\doibase 10.1103/PhysRevB.85.064517}
	{\bibfield  {journal} {\bibinfo  {journal} {Phys. Rev. B}\ }\textbf {\bibinfo
			{volume} {85}},\ \bibinfo {pages} {064517} (\bibinfo {year}
		{2012})}\BibitemShut {NoStop}%
	\bibitem [{\citenamefont {Terashima}\ \emph {et~al.}(2015)\citenamefont
		{Terashima}, \citenamefont {Kikugawa}, \citenamefont {Kasahara},
		\citenamefont {Watashige}, \citenamefont {Shibauchi}, \citenamefont
		{Matsuda}, \citenamefont {Wolf}, \citenamefont {B{\"o}hmer}, \citenamefont
		{Hardy}, \citenamefont {Meingast}, \citenamefont {v.~L{\"o}hneysen},\ and\
		\citenamefont {Uji}}]{Terashima2015}%
	\BibitemOpen
	\bibfield  {author} {\bibinfo {author} {\bibfnamefont {T.}~\bibnamefont
			{Terashima}}, \bibinfo {author} {\bibfnamefont {N.}~\bibnamefont {Kikugawa}},
		\bibinfo {author} {\bibfnamefont {S.}~\bibnamefont {Kasahara}}, \bibinfo
		{author} {\bibfnamefont {T.}~\bibnamefont {Watashige}}, \bibinfo {author}
		{\bibfnamefont {T.}~\bibnamefont {Shibauchi}}, \bibinfo {author}
		{\bibfnamefont {Y.}~\bibnamefont {Matsuda}}, \bibinfo {author} {\bibfnamefont
			{T.}~\bibnamefont {Wolf}}, \bibinfo {author} {\bibfnamefont {A.~E.}\
			\bibnamefont {B{\"o}hmer}}, \bibinfo {author} {\bibfnamefont
			{F.}~\bibnamefont {Hardy}}, \bibinfo {author} {\bibfnamefont
			{C.}~\bibnamefont {Meingast}}, \bibinfo {author} {\bibfnamefont
			{H.}~\bibnamefont {v.~L{\"o}hneysen}}, \ and\ \bibinfo {author}
		{\bibfnamefont {S.}~\bibnamefont {Uji}},\ }\href {\doibase
		10.7566/JPSJ.84.063701} {\bibfield  {journal} {\bibinfo  {journal} {J. Phys.
				Soc. Jpn.}\ }\textbf {\bibinfo {volume} {84}},\ \bibinfo {pages} {063701}
		(\bibinfo {year} {2015})}\BibitemShut {NoStop}%
	\bibitem [{\citenamefont {{Kothapalli}}\ \emph {et~al.}(2016)\citenamefont
		{{Kothapalli}}, \citenamefont {{B{\"o}hmer}}, \citenamefont {{Jayasekara}},
		\citenamefont {{Ueland}}, \citenamefont {{Das}}, \citenamefont {{Sapkota}},
		\citenamefont {{Taufour}}, \citenamefont {{Xiao}}, \citenamefont {{Alp}},
		\citenamefont {{Bud'ko}}, \citenamefont {{Canfield}}, \citenamefont
		{{Kreyssig}},\ and\ \citenamefont {{Goldman}}}]{Kothapalli2016}%
	\BibitemOpen
	\bibfield  {author} {\bibinfo {author} {\bibfnamefont {K.}~\bibnamefont
			{{Kothapalli}}}, \bibinfo {author} {\bibfnamefont {A.~E.}\ \bibnamefont
			{{B{\"o}hmer}}}, \bibinfo {author} {\bibfnamefont {W.~T.}\ \bibnamefont
			{{Jayasekara}}}, \bibinfo {author} {\bibfnamefont {B.~G.}\ \bibnamefont
			{{Ueland}}}, \bibinfo {author} {\bibfnamefont {P.}~\bibnamefont {{Das}}},
		\bibinfo {author} {\bibfnamefont {A.}~\bibnamefont {{Sapkota}}}, \bibinfo
		{author} {\bibfnamefont {V.}~\bibnamefont {{Taufour}}}, \bibinfo {author}
		{\bibfnamefont {Y.}~\bibnamefont {{Xiao}}}, \bibinfo {author} {\bibfnamefont
			{E.~E.}\ \bibnamefont {{Alp}}}, \bibinfo {author} {\bibfnamefont {S.~L.}\
			\bibnamefont {{Bud'ko}}}, \bibinfo {author} {\bibfnamefont {P.~C.}\
			\bibnamefont {{Canfield}}}, \bibinfo {author} {\bibfnamefont
			{A.}~\bibnamefont {{Kreyssig}}}, \ and\ \bibinfo {author} {\bibfnamefont
			{A.~I.}\ \bibnamefont {{Goldman}}},\ }\href@noop {} {\bibfield  {journal} {\bibinfo  {journal} {Nat. Commun.}\ }\textbf {\bibinfo {volume} {7}},\ \bibinfo {pages} {12728}
		(\bibinfo {year} {2016})}
	\BibitemShut {NoStop}%
	\bibitem [{\citenamefont {{Wang}}\ \emph {et~al.}(2016)\citenamefont {{Wang}},
		\citenamefont {{Sun}}, \citenamefont {{Cui}}, \citenamefont {{Song}},
		\citenamefont {{Li}}, \citenamefont {{Yu}}, \citenamefont {{Lei}},\ and\
		\citenamefont {{Yu}}}]{Wang-NMR2016}%
	\BibitemOpen
	\bibfield  {author} {\bibinfo {author} {\bibfnamefont {P.}~\bibnamefont
			{{Wang}}}, \bibinfo {author} {\bibfnamefont {S.}~\bibnamefont {{Sun}}},
		\bibinfo {author} {\bibfnamefont {Y.}~\bibnamefont {{Cui}}}, \bibinfo
		{author} {\bibfnamefont {W.}~\bibnamefont {{Song}}}, \bibinfo {author}
		{\bibfnamefont {T.}~\bibnamefont {{Li}}}, \bibinfo {author} {\bibfnamefont
			{R.}~\bibnamefont {{Yu}}}, \bibinfo {author} {\bibfnamefont {H.}~\bibnamefont
			{{Lei}}}, \ and\ \bibinfo {author} {\bibfnamefont {W.}~\bibnamefont {{Yu}}},\
	}\href@noop {} {\bibfield  {journal} {\bibinfo  {journal} {Phys. Rev. Lett.}\ }\textbf {\bibinfo {volume} {117}},\ \bibinfo {pages} {237001}
		(\bibinfo {year} {2016})} \BibitemShut {NoStop}%
	\bibitem [{\citenamefont {Glasbrenner}\ \emph {et~al.}(2015)\citenamefont
		{Glasbrenner}, \citenamefont {Mazin}, \citenamefont {Jeschke}, \citenamefont
		{Hirschfeld}, \citenamefont {Fernandes},\ and\ \citenamefont
		{Valenti}}]{Glasbrenner2015}%
	\BibitemOpen
	\bibfield  {author} {\bibinfo {author} {\bibfnamefont {J.~K.}\ \bibnamefont
			{Glasbrenner}}, \bibinfo {author} {\bibfnamefont {I.~I.}\ \bibnamefont
			{Mazin}}, \bibinfo {author} {\bibfnamefont {H.~O.}\ \bibnamefont {Jeschke}},
		\bibinfo {author} {\bibfnamefont {P.~J.}\ \bibnamefont {Hirschfeld}},
		\bibinfo {author} {\bibfnamefont {R.~M.}\ \bibnamefont {Fernandes}}, \ and\
		\bibinfo {author} {\bibfnamefont {R.}~\bibnamefont {Valenti}},\ }\href
	{http://dx.doi.org/10.1038/nphys3434} {\bibfield  {journal} {\bibinfo
			{journal} {Nat. Phys.}\ }\textbf {\bibinfo {volume} {11}},\ \bibinfo {pages}
		{953} (\bibinfo {year} {2015})}\BibitemShut {NoStop}%
	\bibitem [{\citenamefont {Wang}\ \emph
		{et~al.}(2015{\natexlab{b}})\citenamefont {Wang}, \citenamefont {Kivelson},\
		and\ \citenamefont {Lee}}]{Wang2015}%
	\BibitemOpen
	\bibfield  {author} {\bibinfo {author} {\bibfnamefont {F.}~\bibnamefont
			{Wang}}, \bibinfo {author} {\bibfnamefont {S.~A.}\ \bibnamefont {Kivelson}},
		\ and\ \bibinfo {author} {\bibfnamefont {D.-H.}\ \bibnamefont {Lee}},\
	}\href@noop {} {\bibfield  {journal} {\bibinfo  {journal} {Nat. Phys.}\
		}\textbf {\bibinfo {volume} {11}},\ \bibinfo {pages} {959} (\bibinfo {year}
		{2015}{\natexlab{b}})}\BibitemShut {NoStop}%
	\bibitem [{\citenamefont {Yu}\ and\ \citenamefont {Si}(2015)}]{Yu2015}%
	\BibitemOpen
	\bibfield  {author} {\bibinfo {author} {\bibfnamefont {R.}~\bibnamefont
			{Yu}}\ and\ \bibinfo {author} {\bibfnamefont {Q.}~\bibnamefont {Si}},\ }\href
	{\doibase 10.1103/PhysRevLett.115.116401} {\bibfield  {journal} {\bibinfo
			{journal} {Phys. Rev. Lett.}\ }\textbf {\bibinfo {volume} {115}},\ \bibinfo
		{pages} {116401} (\bibinfo {year} {2015})}\BibitemShut {NoStop}%
	\bibitem [{\citenamefont {Wang}\ \emph {et~al.}(2016)\citenamefont {Wang},
		\citenamefont {Hu},\ and\ \citenamefont {Nevidomskyy}}]{ZWang2016}%
	\BibitemOpen
	\bibfield  {author} {\bibinfo {author} {\bibfnamefont {Z.}~\bibnamefont
			{Wang}}, \bibinfo {author} {\bibfnamefont {W.~J.}\ \bibnamefont {Hu}}, \ and\
		\bibinfo {author} {\bibfnamefont {A.~H.}\ \bibnamefont {Nevidomskyy}},\
	}\href@noop {} {\bibfield  {journal} {\bibinfo  {journal} {Phys. Rev. Lett.}\
		}\textbf {\bibinfo {volume} {116}},\ \bibinfo {pages} {247203} (\bibinfo
		{year} {2016})}\BibitemShut {NoStop}%
	\bibitem [{\citenamefont {Fang}\ \emph {et~al.}(2008)\citenamefont {Fang},
		\citenamefont {Yao}, \citenamefont {Tsai}, \citenamefont {Hu},\ and\
		\citenamefont {Kivelson}}]{Fang2008}%
	\BibitemOpen
	\bibfield  {author} {\bibinfo {author} {\bibfnamefont {C.}~\bibnamefont
			{Fang}}, \bibinfo {author} {\bibfnamefont {H.}~\bibnamefont {Yao}}, \bibinfo
		{author} {\bibfnamefont {W.-F.}\ \bibnamefont {Tsai}}, \bibinfo {author}
		{\bibfnamefont {J.~P.}\ \bibnamefont {Hu}}, \ and\ \bibinfo {author}
		{\bibfnamefont {S.~A.}\ \bibnamefont {Kivelson}},\ }\href {\doibase
		10.1103/PhysRevB.77.224509} {\bibfield  {journal} {\bibinfo  {journal} {Phys.
				Rev. B}\ }\textbf {\bibinfo {volume} {77}},\ \bibinfo {pages} {224509}
		(\bibinfo {year} {2008})}\BibitemShut {NoStop}%
	\bibitem [{\citenamefont {Wysocki}\ \emph {et~al.}(2011)\citenamefont
		{Wysocki}, \citenamefont {Belashchenko},\ and\ \citenamefont
		{Antropov}}]{Wysocki2011}%
	\BibitemOpen
	\bibfield  {author} {\bibinfo {author} {\bibfnamefont {A.~L.}\ \bibnamefont
			{Wysocki}}, \bibinfo {author} {\bibfnamefont {K.~D.}\ \bibnamefont
			{Belashchenko}}, \ and\ \bibinfo {author} {\bibfnamefont {V.~P.}\
			\bibnamefont {Antropov}},\ }\href@noop {} {\bibfield  {journal} {\bibinfo
			{journal} {Nat. Phys.}\ }\textbf {\bibinfo {volume} {7}},\ \bibinfo {pages}
		{485} (\bibinfo {year} {2011})}\BibitemShut {NoStop}%
	\bibitem [{\citenamefont {Yu}\ \emph {et~al.}(2012)\citenamefont {Yu},
		\citenamefont {Wang}, \citenamefont {Goswami}, \citenamefont {Nevidomskyy},
		\citenamefont {Si},\ and\ \citenamefont {Abrahams}}]{Yu2012}%
	\BibitemOpen
	\bibfield  {author} {\bibinfo {author} {\bibfnamefont {R.}~\bibnamefont
			{Yu}}, \bibinfo {author} {\bibfnamefont {Z.}~\bibnamefont {Wang}}, \bibinfo
		{author} {\bibfnamefont {P.}~\bibnamefont {Goswami}}, \bibinfo {author}
		{\bibfnamefont {A.~H.}\ \bibnamefont {Nevidomskyy}}, \bibinfo {author}
		{\bibfnamefont {Q.}~\bibnamefont {Si}}, \ and\ \bibinfo {author}
		{\bibfnamefont {E.}~\bibnamefont {Abrahams}},\ }\href {\doibase
		10.1103/PhysRevB.86.085148} {\bibfield  {journal} {\bibinfo  {journal} {Phys.
				Rev. B}\ }\textbf {\bibinfo {volume} {86}},\ \bibinfo {pages} {085148}
		(\bibinfo {year} {2012})}\BibitemShut {NoStop}%
	\bibitem [{\citenamefont {Lai}\ \emph {et~al.}(2016)\citenamefont {Lai},
		\citenamefont {Hu}, \citenamefont {Yu},\ and\ \citenamefont {Si}}]{Lai2016}%
	\BibitemOpen
	\bibfield  {author} {\bibinfo {author} {\bibfnamefont {H.-H.}\ \bibnamefont
			{Lai}}, \bibinfo {author} {\bibfnamefont {W.-J.}\ \bibnamefont {Hu}},
		\bibinfo {author} {\bibfnamefont {R.}~\bibnamefont {Yu}}, \ and\ \bibinfo
		{author} {\bibfnamefont {Q.}~\bibnamefont {Si}},\ }\href@noop {} {\bibfield  {journal} {\bibinfo  {journal} {Phys.
				Rev. Lett.}\ }\textbf {\bibinfo {volume} {118}},\ \bibinfo {pages} {176401}
		(\bibinfo {year} {2012})}\BibitemShut {NoStop}%
	\bibitem [{\citenamefont {Hu}\ \emph {et~al.}(2016)\citenamefont {Hu},
		\citenamefont {Lai}, \citenamefont {Gong}, \citenamefont {Yu}, \citenamefont
		{Nevidomskyy},\ and\ \citenamefont {Si}}]{Hu2016}%
	\BibitemOpen
	\bibfield  {author} {\bibinfo {author} {\bibfnamefont {W.-J.}\ \bibnamefont
			{Hu}}, \bibinfo {author} {\bibfnamefont {H.-H.}\ \bibnamefont {Lai}},
		\bibinfo {author} {\bibfnamefont {S.-S.}\ \bibnamefont {Gong}}, \bibinfo
		{author} {\bibfnamefont {R.}~\bibnamefont {Yu}}, \bibinfo {author}
		{\bibfnamefont {A.~H.}\ \bibnamefont {Nevidomskyy}}, \ and\ \bibinfo {author}
		{\bibfnamefont {Q.}~\bibnamefont {Si}},\ }\href@noop {} {\bibfield  {journal}
		{\bibinfo  {journal} {arXiv:1606.01235}\ } (\bibinfo {year}
		{2016})}\BibitemShut {NoStop}%
	\bibitem [{\citenamefont {Lipscombe}\ \emph {et~al.}(2011)\citenamefont
		{Lipscombe}, \citenamefont {Chen}, \citenamefont {Fang}, \citenamefont
		{Perring}, \citenamefont {Abernathy}, \citenamefont {Christianson},
		\citenamefont {Egami}, \citenamefont {Wang}, \citenamefont {Hu},\ and\
		\citenamefont {Dai}}]{Lipscombe2011}%
	\BibitemOpen
	\bibfield  {author} {\bibinfo {author} {\bibfnamefont {O.~J.}\ \bibnamefont
			{Lipscombe}}, \bibinfo {author} {\bibfnamefont {G.~F.}\ \bibnamefont {Chen}},
		\bibinfo {author} {\bibfnamefont {C.}~\bibnamefont {Fang}}, \bibinfo {author}
		{\bibfnamefont {T.~G.}\ \bibnamefont {Perring}}, \bibinfo {author}
		{\bibfnamefont {D.~L.}\ \bibnamefont {Abernathy}}, \bibinfo {author}
		{\bibfnamefont {A.~D.}\ \bibnamefont {Christianson}}, \bibinfo {author}
		{\bibfnamefont {T.}~\bibnamefont {Egami}}, \bibinfo {author} {\bibfnamefont
			{N.}~\bibnamefont {Wang}}, \bibinfo {author} {\bibfnamefont {J.}~\bibnamefont
			{Hu}}, \ and\ \bibinfo {author} {\bibfnamefont {P.}~\bibnamefont {Dai}},\
	}\href {\doibase 10.1103/PhysRevLett.106.057004} {\bibfield  {journal}
		{\bibinfo  {journal} {Phys. Rev. Lett.}\ }\textbf {\bibinfo {volume} {106}},\
		\bibinfo {pages} {057004} (\bibinfo {year} {2011})}\BibitemShut {NoStop}%
	\bibitem [{\citenamefont {Zhao}\ \emph {et~al.}(2009)\citenamefont {Zhao},
		\citenamefont {Adroja}, \citenamefont {Yao}, \citenamefont {Bewley},
		\citenamefont {Li}, \citenamefont {Wang}, \citenamefont {Wu}, \citenamefont
		{Chen}, \citenamefont {Hu},\ and\ \citenamefont {Dai}}]{Zhao2009}%
	\BibitemOpen
	\bibfield  {author} {\bibinfo {author} {\bibfnamefont {J.}~\bibnamefont
			{Zhao}}, \bibinfo {author} {\bibfnamefont {D.~T.}\ \bibnamefont {Adroja}},
		\bibinfo {author} {\bibfnamefont {D.-X.}\ \bibnamefont {Yao}}, \bibinfo
		{author} {\bibfnamefont {R.}~\bibnamefont {Bewley}}, \bibinfo {author}
		{\bibfnamefont {S.}~\bibnamefont {Li}}, \bibinfo {author} {\bibfnamefont
			{X.~F.}\ \bibnamefont {Wang}}, \bibinfo {author} {\bibfnamefont
			{G.}~\bibnamefont {Wu}}, \bibinfo {author} {\bibfnamefont {X.~H.}\
			\bibnamefont {Chen}}, \bibinfo {author} {\bibfnamefont {J.}~\bibnamefont
			{Hu}}, \ and\ \bibinfo {author} {\bibfnamefont {P.}~\bibnamefont {Dai}},\
	}\href {http://dx.doi.org/10.1038/nphys1336} {\bibfield  {journal} {\bibinfo
			{journal} {Nat. Phys.}\ }\textbf {\bibinfo {volume} {5}},\ \bibinfo {pages}
		{555} (\bibinfo {year} {2009})}\BibitemShut {NoStop}%
	\bibitem [{\citenamefont {Harriger}\ \emph {et~al.}(2011)\citenamefont
		{Harriger}, \citenamefont {Luo}, \citenamefont {Liu}, \citenamefont {Frost},
		\citenamefont {Hu}, \citenamefont {Norman},\ and\ \citenamefont
		{Dai}}]{Harriger2011}%
	\BibitemOpen
	\bibfield  {author} {\bibinfo {author} {\bibfnamefont {L.~W.}\ \bibnamefont
			{Harriger}}, \bibinfo {author} {\bibfnamefont {H.~Q.}\ \bibnamefont {Luo}},
		\bibinfo {author} {\bibfnamefont {M.~S.}\ \bibnamefont {Liu}}, \bibinfo
		{author} {\bibfnamefont {C.}~\bibnamefont {Frost}}, \bibinfo {author}
		{\bibfnamefont {J.~P.}\ \bibnamefont {Hu}}, \bibinfo {author} {\bibfnamefont
			{M.~R.}\ \bibnamefont {Norman}}, \ and\ \bibinfo {author} {\bibfnamefont
			{P.}~\bibnamefont {Dai}},\ }\href {\doibase 10.1103/PhysRevB.84.054544}
	{\bibfield  {journal} {\bibinfo  {journal} {Phys. Rev. B}\ }\textbf {\bibinfo
			{volume} {84}},\ \bibinfo {pages} {054544} (\bibinfo {year}
		{2011})}\BibitemShut {NoStop}%
	\bibitem [{\citenamefont {Papanicolaou}(1988)}]{Papanicolaou1988}%
	\BibitemOpen
	\bibfield  {author} {\bibinfo {author} {\bibfnamefont {N.}~\bibnamefont
			{Papanicolaou}},\ }\href {\doibase
		http://dx.doi.org/10.1016/0550-3213(88)90073-9} {\bibfield  {journal}
		{\bibinfo  {journal} {Nucl. Phys. B}\ }\textbf {\bibinfo {volume} {305}},\
		\bibinfo {pages} {367 } (\bibinfo {year} {1988})}\BibitemShut {NoStop}%
	\bibitem [{\citenamefont {Tsunetsugu}\ and\ \citenamefont
		{Arikawa}(2007)}]{Tsunetsugu2006}%
	\BibitemOpen
	\bibfield  {author} {\bibinfo {author} {\bibfnamefont {H.}~\bibnamefont
			{Tsunetsugu}}\ and\ \bibinfo {author} {\bibfnamefont {M.}~\bibnamefont
			{Arikawa}},\ }\href@noop {} {\bibfield  {journal} {\bibinfo  {journal} {J.
				Phys.: Condensed Matter}\ }\textbf {\bibinfo {volume} {19}},\ \bibinfo
		{pages} {145248} (\bibinfo {year} {2007})}\BibitemShut {NoStop}%
	\bibitem [{\citenamefont {L\"auchli}\ \emph {et~al.}(2006)\citenamefont
		{L\"auchli}, \citenamefont {Mila},\ and\ \citenamefont {Penc}}]{Lauchli2006}%
	\BibitemOpen
	\bibfield  {author} {\bibinfo {author} {\bibfnamefont {A.}~\bibnamefont
			{L\"auchli}}, \bibinfo {author} {\bibfnamefont {F.}~\bibnamefont {Mila}}, \
		and\ \bibinfo {author} {\bibfnamefont {K.}~\bibnamefont {Penc}},\ }\href
	{\doibase 10.1103/PhysRevLett.97.087205} {\bibfield  {journal} {\bibinfo
			{journal} {Phys. Rev. Lett.}\ }\textbf {\bibinfo {volume} {97}},\ \bibinfo
		{pages} {087205} (\bibinfo {year} {2006})}\BibitemShut {NoStop}%
	\bibitem [{\citenamefont {Muniz}\ \emph {et~al.}(2014)\citenamefont {Muniz},
		\citenamefont {Kato},\ and\ \citenamefont {Batista}}]{Muniz2014}%
	\BibitemOpen
	\bibfield  {author} {\bibinfo {author} {\bibfnamefont {R.~A.}\ \bibnamefont
			{Muniz}}, \bibinfo {author} {\bibfnamefont {Y.}~\bibnamefont {Kato}}, \ and\
		\bibinfo {author} {\bibfnamefont {C.~D.}\ \bibnamefont {Batista}},\ }\href
	{\doibase 10.1093/ptep/ptu109} {\bibfield  {journal} {\bibinfo  {journal}
			{Prog. Theor. Exp. Phys.}\ }\textbf {\bibinfo {volume} {2014}},\ \bibinfo
		{pages} {083I01} (\bibinfo {year} {2014})}\BibitemShut {NoStop}%
	\bibitem [{\citenamefont {Chaloupka}\ and\ \citenamefont
		{Khaliullin}(2013)}]{Chaloupka2013}%
	\BibitemOpen
	\bibfield  {author} {\bibinfo {author} {\bibfnamefont {J.}~\bibnamefont
			{Chaloupka}}\ and\ \bibinfo {author} {\bibfnamefont {G.}~\bibnamefont
			{Khaliullin}},\ }\href {\doibase 10.1103/PhysRevLett.110.207205} {\bibfield
		{journal} {\bibinfo  {journal} {Phys. Rev. Lett.}\ }\textbf {\bibinfo
			{volume} {110}},\ \bibinfo {pages} {207205} (\bibinfo {year}
		{2013})}\BibitemShut {NoStop}%
	\bibitem [{\citenamefont {Bilbao~Ergueta}\ and\ \citenamefont
		{Nevidomskyy}(2015)}]{Ergueta2015}%
	\BibitemOpen
	\bibfield  {author} {\bibinfo {author} {\bibfnamefont {P.}~\bibnamefont
			{Bilbao~Ergueta}}\ and\ \bibinfo {author} {\bibfnamefont {A.~H.}\
			\bibnamefont {Nevidomskyy}},\ }\href {\doibase 10.1103/PhysRevB.92.165102}
	{\bibfield  {journal} {\bibinfo  {journal} {Phys. Rev. B}\ }\textbf {\bibinfo
			{volume} {92}},\ \bibinfo {pages} {165102} (\bibinfo {year}
		{2015})}\BibitemShut {NoStop}%
	\bibitem [{\citenamefont {White}(1992)}]{White1992}%
	\BibitemOpen
	\bibfield  {author} {\bibinfo {author} {\bibfnamefont {S.~R.}\ \bibnamefont
			{White}},\ }\href {\doibase 10.1103/PhysRevLett.69.2863} {\bibfield
		{journal} {\bibinfo  {journal} {Phys. Rev. Lett.}\ }\textbf {\bibinfo
			{volume} {69}},\ \bibinfo {pages} {2863} (\bibinfo {year}
		{1992})}\BibitemShut {NoStop}%
	\bibitem [{\citenamefont {McCulloch}\ and\ \citenamefont
		{Gul{\'a}csi}(2002)}]{mcculloch2002}%
	\BibitemOpen
	\bibfield  {author} {\bibinfo {author} {\bibfnamefont {I.}~\bibnamefont
			{McCulloch}}\ and\ \bibinfo {author} {\bibfnamefont {M.}~\bibnamefont
			{Gul{\'a}csi}},\ }\href {http://iopscience.iop.org/0295-5075/57/6/852}
	{\bibfield  {journal} {\bibinfo  {journal} {Europhys. Lett.}\ }\textbf
		{\bibinfo {volume} {57}},\ \bibinfo {pages} {852} (\bibinfo {year}
		{2002})}\BibitemShut {NoStop}%
	\bibitem [{\citenamefont {Gong}\ \emph
		{et~al.}(2014{\natexlab{a}})\citenamefont {Gong}, \citenamefont {Zhu},
		\citenamefont {Sheng}, \citenamefont {Motrunich},\ and\ \citenamefont
		{Fisher}}]{gong2014square}%
	\BibitemOpen
	\bibfield  {author} {\bibinfo {author} {\bibfnamefont {S.-S.}\ \bibnamefont
			{Gong}}, \bibinfo {author} {\bibfnamefont {W.}~\bibnamefont {Zhu}}, \bibinfo
		{author} {\bibfnamefont {D.~N.}\ \bibnamefont {Sheng}}, \bibinfo {author}
		{\bibfnamefont {O.~I.}\ \bibnamefont {Motrunich}}, \ and\ \bibinfo {author}
		{\bibfnamefont {M.~P.~A.}\ \bibnamefont {Fisher}},\ }\href {\doibase
		10.1103/PhysRevLett.113.027201} {\bibfield  {journal} {\bibinfo  {journal}
			{Phys. Rev. Lett.}\ }\textbf {\bibinfo {volume} {113}},\ \bibinfo {pages}
		{027201} (\bibinfo {year} {2014}{\natexlab{a}})}\BibitemShut {NoStop}%
	\bibitem [{\citenamefont {Gong}\ \emph
		{et~al.}(2014{\natexlab{b}})\citenamefont {Gong}, \citenamefont {Zhu},\ and\
		\citenamefont {Sheng}}]{gong2014kagome}%
	\BibitemOpen
	\bibfield  {author} {\bibinfo {author} {\bibfnamefont {S.-S.}\ \bibnamefont
			{Gong}}, \bibinfo {author} {\bibfnamefont {W.}~\bibnamefont {Zhu}}, \ and\
		\bibinfo {author} {\bibfnamefont {D.}~\bibnamefont {Sheng}},\ }\href
	{http://www.nature.com/srep/2014/140910/srep06317/full/srep06317.html}
	{\bibfield  {journal} {\bibinfo  {journal} {Sci. Rep.}\ }\textbf {\bibinfo
			{volume} {4}},\ \bibinfo {pages} {6317} (\bibinfo {year}
		{2014}{\natexlab{b}})}\BibitemShut {NoStop}%
	\bibitem [{Note1()}]{Note1}%
	\BibitemOpen
	\bibinfo {note} {$L$ represents the size of y-direction which has periodic
		boundary condition}\BibitemShut {NoStop}%
	\bibitem [{\citenamefont {White}\ and\ \citenamefont
		{Chernyshev}(2007)}]{white2007}%
	\BibitemOpen
	\bibfield  {author} {\bibinfo {author} {\bibfnamefont {S.~R.}\ \bibnamefont
			{White}}\ and\ \bibinfo {author} {\bibfnamefont {A.~L.}\ \bibnamefont
			{Chernyshev}},\ }\href {\doibase 10.1103/PhysRevLett.99.127004} {\bibfield
		{journal} {\bibinfo  {journal} {Phys. Rev. Lett.}\ }\textbf {\bibinfo
			{volume} {99}},\ \bibinfo {pages} {127004} (\bibinfo {year}
		{2007})}\BibitemShut {NoStop}%
	\bibitem [{\citenamefont {Yan}\ \emph {et~al.}(2011)\citenamefont {Yan},
		\citenamefont {Huse},\ and\ \citenamefont {White}}]{Yan2011}%
	\BibitemOpen
	\bibfield  {author} {\bibinfo {author} {\bibfnamefont {S.}~\bibnamefont
			{Yan}}, \bibinfo {author} {\bibfnamefont {D.~A.}\ \bibnamefont {Huse}}, \
		and\ \bibinfo {author} {\bibfnamefont {S.~R.}\ \bibnamefont {White}},\ }\href
	{\doibase 10.1126/science.1201080} {\bibfield  {journal} {\bibinfo  {journal}
			{Science}\ }\textbf {\bibinfo {volume} {332}},\ \bibinfo {pages} {1173}
		(\bibinfo {year} {2011})}\BibitemShut {NoStop}%
	\bibitem [{\citenamefont {Gong}\ \emph {et~al.}(2015)\citenamefont {Gong},
		\citenamefont {Zhu},\ and\ \citenamefont {Sheng}}]{Gong2015_honeycomb}%
	\BibitemOpen
	\bibfield  {author} {\bibinfo {author} {\bibfnamefont {S.-S.}\ \bibnamefont
			{Gong}}, \bibinfo {author} {\bibfnamefont {W.}~\bibnamefont {Zhu}}, \ and\
		\bibinfo {author} {\bibfnamefont {D.~N.}\ \bibnamefont {Sheng}},\ }\href
	{\doibase 10.1103/PhysRevB.92.195110} {\bibfield  {journal} {\bibinfo
			{journal} {Phys. Rev. B}\ }\textbf {\bibinfo {volume} {92}},\ \bibinfo
		{pages} {195110} (\bibinfo {year} {2015})}\BibitemShut {NoStop}%
	\bibitem [{\citenamefont {Moon}\ and\ \citenamefont {Choi}(2010)}]{Moon2010}%
	\BibitemOpen
	\bibfield  {author} {\bibinfo {author} {\bibfnamefont {C.-Y.}\ \bibnamefont
			{Moon}}\ and\ \bibinfo {author} {\bibfnamefont {H.~J.}\ \bibnamefont
			{Choi}},\ }\href@noop {} {\bibfield  {journal} {\bibinfo  {journal} {Phys.
				Rev. Lett.}\ }\textbf {\bibinfo {volume} {104}},\ \bibinfo {pages} {057003}
		(\bibinfo {year} {2010})}\BibitemShut {NoStop}%
	\bibitem [{\citenamefont {McQueen}\ \emph
		{et~al.}(2009{\natexlab{b}})\citenamefont {McQueen}, \citenamefont
		{Williams}, \citenamefont {Stephens}, \citenamefont {Tao}, \citenamefont
		{Zhu}, \citenamefont {Ksenofontov}, \citenamefont {Casper}, \citenamefont
		{Felser},\ and\ \citenamefont {Cava}}]{McQueen2009b}%
	\BibitemOpen
	\bibfield  {author} {\bibinfo {author} {\bibfnamefont {T.~M.}\ \bibnamefont
			{McQueen}}, \bibinfo {author} {\bibfnamefont {A.~J.}\ \bibnamefont
			{Williams}}, \bibinfo {author} {\bibfnamefont {P.~W.}\ \bibnamefont
			{Stephens}}, \bibinfo {author} {\bibfnamefont {J.}~\bibnamefont {Tao}},
		\bibinfo {author} {\bibfnamefont {Y.}~\bibnamefont {Zhu}}, \bibinfo {author}
		{\bibfnamefont {V.}~\bibnamefont {Ksenofontov}}, \bibinfo {author}
		{\bibfnamefont {F.}~\bibnamefont {Casper}}, \bibinfo {author} {\bibfnamefont
			{C.}~\bibnamefont {Felser}}, \ and\ \bibinfo {author} {\bibfnamefont {R.~J.}\
			\bibnamefont {Cava}},\ }\href@noop {} {\bibfield  {journal} {\bibinfo
			{journal} {Phys. Rev. Lett.}\ }\textbf {\bibinfo {volume} {103}},\ \bibinfo
		{pages} {057002} (\bibinfo {year} {2009}{\natexlab{b}})}\BibitemShut
	{NoStop}%
	\bibitem [{\citenamefont {Louca}\ \emph {et~al.}(2010)\citenamefont {Louca},
		\citenamefont {Horigane}, \citenamefont {Llobet}, \citenamefont {Arita},
		\citenamefont {Ji}, \citenamefont {Katayama}, \citenamefont {Konbu},
		\citenamefont {Nakamura}, \citenamefont {Koo}, \citenamefont {Tong},\ and\
		\citenamefont {Yamada}}]{Louca2010}%
	\BibitemOpen
	\bibfield  {author} {\bibinfo {author} {\bibfnamefont {D.}~\bibnamefont
			{Louca}}, \bibinfo {author} {\bibfnamefont {K.}~\bibnamefont {Horigane}},
		\bibinfo {author} {\bibfnamefont {A.}~\bibnamefont {Llobet}}, \bibinfo
		{author} {\bibfnamefont {R.}~\bibnamefont {Arita}}, \bibinfo {author}
		{\bibfnamefont {S.}~\bibnamefont {Ji}}, \bibinfo {author} {\bibfnamefont
			{N.}~\bibnamefont {Katayama}}, \bibinfo {author} {\bibfnamefont
			{S.}~\bibnamefont {Konbu}}, \bibinfo {author} {\bibfnamefont
			{K.}~\bibnamefont {Nakamura}}, \bibinfo {author} {\bibfnamefont {T.~Y.}\
			\bibnamefont {Koo}}, \bibinfo {author} {\bibfnamefont {P.}~\bibnamefont
			{Tong}}, \ and\ \bibinfo {author} {\bibfnamefont {K.}~\bibnamefont
			{Yamada}},\ }\href@noop {} {\bibfield  {journal} {\bibinfo  {journal} {Phys.
				Rev. B}\ }\textbf {\bibinfo {volume} {81}},\ \bibinfo {pages} {134524}
		(\bibinfo {year} {2010})}\BibitemShut {NoStop}%
	\bibitem [{\citenamefont {Xu}\ \emph {et~al.}(2012)\citenamefont {Xu},
		\citenamefont {Wen}, \citenamefont {Zhao}, \citenamefont {Matsuda},
		\citenamefont {Ku}, \citenamefont {Liu}, \citenamefont {Gu}, \citenamefont
		{Lee}, \citenamefont {Birgeneau}, \citenamefont {Tranquada},\ and\
		\citenamefont {Xu}}]{Xu2012}%
	\BibitemOpen
	\bibfield  {author} {\bibinfo {author} {\bibfnamefont {Z.}~\bibnamefont
			{Xu}}, \bibinfo {author} {\bibfnamefont {J.}~\bibnamefont {Wen}}, \bibinfo
		{author} {\bibfnamefont {Y.}~\bibnamefont {Zhao}}, \bibinfo {author}
		{\bibfnamefont {M.}~\bibnamefont {Matsuda}}, \bibinfo {author} {\bibfnamefont
			{W.}~\bibnamefont {Ku}}, \bibinfo {author} {\bibfnamefont {X.}~\bibnamefont
			{Liu}}, \bibinfo {author} {\bibfnamefont {G.}~\bibnamefont {Gu}}, \bibinfo
		{author} {\bibfnamefont {D.~H.}\ \bibnamefont {Lee}}, \bibinfo {author}
		{\bibfnamefont {R.~J.}\ \bibnamefont {Birgeneau}}, \bibinfo {author}
		{\bibfnamefont {J.~M.}\ \bibnamefont {Tranquada}}, \ and\ \bibinfo {author}
		{\bibfnamefont {G.}~\bibnamefont {Xu}},\ }\href@noop {} {\bibfield  {journal}
		{\bibinfo  {journal} {Phys. Rev. Lett.}\ }\textbf {\bibinfo {volume} {109}},\
		\bibinfo {pages} {227002} (\bibinfo {year} {2012})}\BibitemShut {NoStop}%
	\bibitem [{\citenamefont {Tsyrulin}\ \emph {et~al.}(2012)\citenamefont
		{Tsyrulin}, \citenamefont {Viennois}, \citenamefont {Giannini}, \citenamefont
		{Boehm}, \citenamefont {Jimenez-Ruiz}, \citenamefont {Omrani}, \citenamefont
		{Piazza},\ and\ \citenamefont {Ronnow}}]{Tsyrulin2012}%
	\BibitemOpen
	\bibfield  {author} {\bibinfo {author} {\bibfnamefont {N.}~\bibnamefont
			{Tsyrulin}}, \bibinfo {author} {\bibfnamefont {R.}~\bibnamefont {Viennois}},
		\bibinfo {author} {\bibfnamefont {E.}~\bibnamefont {Giannini}}, \bibinfo
		{author} {\bibfnamefont {M.}~\bibnamefont {Boehm}}, \bibinfo {author}
		{\bibfnamefont {M.}~\bibnamefont {Jimenez-Ruiz}}, \bibinfo {author}
		{\bibfnamefont {A.~A.}\ \bibnamefont {Omrani}}, \bibinfo {author}
		{\bibfnamefont {B.~D.}\ \bibnamefont {Piazza}}, \ and\ \bibinfo {author}
		{\bibfnamefont {H.~M.}\ \bibnamefont {Ronnow}},\ }\href@noop {} {\bibfield
		{journal} {\bibinfo  {journal} {New J. Phys.}\ }\textbf {\bibinfo {volume}
			{14}},\ \bibinfo {pages} {073025} (\bibinfo {year} {2012})}\BibitemShut
	{NoStop}%
	\bibitem [{\citenamefont {Xu}\ \emph {et~al.}(2014)\citenamefont {Xu},
		\citenamefont {Wen}, \citenamefont {Schneeloch}, \citenamefont
		{Christianson}, \citenamefont {Birgeneau}, \citenamefont {Gu}, \citenamefont
		{Tranquada},\ and\ \citenamefont {Xu}}]{Xu2014}%
	\BibitemOpen
	\bibfield  {author} {\bibinfo {author} {\bibfnamefont {Z.}~\bibnamefont
			{Xu}}, \bibinfo {author} {\bibfnamefont {J.}~\bibnamefont {Wen}}, \bibinfo
		{author} {\bibfnamefont {J.}~\bibnamefont {Schneeloch}}, \bibinfo {author}
		{\bibfnamefont {A.~D.}\ \bibnamefont {Christianson}}, \bibinfo {author}
		{\bibfnamefont {R.~J.}\ \bibnamefont {Birgeneau}}, \bibinfo {author}
		{\bibfnamefont {G.}~\bibnamefont {Gu}}, \bibinfo {author} {\bibfnamefont
			{J.~M.}\ \bibnamefont {Tranquada}}, \ and\ \bibinfo {author} {\bibfnamefont
			{G.}~\bibnamefont {Xu}},\ }\href@noop {} {\bibfield  {journal} {\bibinfo
			{journal} {Phys. Rev. B}\ }\textbf {\bibinfo {volume} {89}},\ \bibinfo
		{pages} {174517} (\bibinfo {year} {2014})}\BibitemShut {NoStop}%
	\bibitem [{\citenamefont {Yu}\ and\ \citenamefont {Si}(2012)}]{Yu2012-OSMP}%
	\BibitemOpen
	\bibfield  {author} {\bibinfo {author} {\bibfnamefont {R.}~\bibnamefont
			{Yu}}\ and\ \bibinfo {author} {\bibfnamefont {Q.}~\bibnamefont {Si}},\ }\href
	{\doibase 10.1103/PhysRevB.86.085104} {\bibfield  {journal} {\bibinfo
			{journal} {Phys. Rev. B}\ }\textbf {\bibinfo {volume} {86}},\ \bibinfo
		{pages} {085104} (\bibinfo {year} {2012})}\BibitemShut {NoStop}%
	\bibitem [{\citenamefont {Bascones}\ \emph {et~al.}(2012)\citenamefont
		{Bascones}, \citenamefont {Valenzuela},\ and\ \citenamefont
		{Calder\'on}}]{Bascones2012}%
	\BibitemOpen
	\bibfield  {author} {\bibinfo {author} {\bibfnamefont {E.}~\bibnamefont
			{Bascones}}, \bibinfo {author} {\bibfnamefont {B.}~\bibnamefont
			{Valenzuela}}, \ and\ \bibinfo {author} {\bibfnamefont {M.~J.}\ \bibnamefont
			{Calder\'on}},\ }\href {\doibase 10.1103/PhysRevB.86.174508} {\bibfield
		{journal} {\bibinfo  {journal} {Phys. Rev. B}\ }\textbf {\bibinfo {volume}
			{86}},\ \bibinfo {pages} {174508} (\bibinfo {year} {2012})}\BibitemShut
	{NoStop}%
	\bibitem [{\citenamefont {Yu}\ and\ \citenamefont {Si}(2013)}]{Yu2013}%
	\BibitemOpen
	\bibfield  {author} {\bibinfo {author} {\bibfnamefont {R.}~\bibnamefont
			{Yu}}\ and\ \bibinfo {author} {\bibfnamefont {Q.}~\bibnamefont {Si}},\ }\href
	{\doibase 10.1103/PhysRevLett.110.146402} {\bibfield  {journal} {\bibinfo
			{journal} {Phys. Rev. Lett.}\ }\textbf {\bibinfo {volume} {110}},\ \bibinfo
		{pages} {146402} (\bibinfo {year} {2013})}\BibitemShut {NoStop}%
	\bibitem [{\citenamefont {de' Medici}\ \emph {et~al.}(2014)\citenamefont {de'
			Medici}, \citenamefont {Giovannetti},\ and\ \citenamefont
		{Capone}}]{Medici2014}%
	\BibitemOpen
	\bibfield  {author} {\bibinfo {author} {\bibfnamefont {L.}~\bibnamefont {de'
				Medici}}, \bibinfo {author} {\bibfnamefont {G.}~\bibnamefont {Giovannetti}},
		\ and\ \bibinfo {author} {\bibfnamefont {M.}~\bibnamefont {Capone}},\ }\href
	{\doibase 10.1103/PhysRevLett.112.177001} {\bibfield  {journal} {\bibinfo
			{journal} {Phys. Rev. Lett.}\ }\textbf {\bibinfo {volume} {112}},\ \bibinfo
		{pages} {177001} (\bibinfo {year} {2014})}\BibitemShut {NoStop}%
	\bibitem [{\citenamefont {Seo}\ \emph {et~al.}(2008)\citenamefont {Seo},
		\citenamefont {Bernevig},\ and\ \citenamefont {Hu}}]{Seo2008}%
	\BibitemOpen
	\bibfield  {author} {\bibinfo {author} {\bibfnamefont {K.}~\bibnamefont
			{Seo}}, \bibinfo {author} {\bibfnamefont {B.~A.}\ \bibnamefont {Bernevig}}, \
		and\ \bibinfo {author} {\bibfnamefont {J.}~\bibnamefont {Hu}},\ }\href
	{\doibase 10.1103/PhysRevLett.101.206404} {\bibfield  {journal} {\bibinfo
			{journal} {Phys. Rev. Lett.}\ }\textbf {\bibinfo {volume} {101}},\ \bibinfo
		{pages} {206404} (\bibinfo {year} {2008})}\BibitemShut {NoStop}%
	\bibitem [{\citenamefont {Chen}\ \emph {et~al.}(2009)\citenamefont {Chen},
		\citenamefont {Yang}, \citenamefont {Zhou},\ and\ \citenamefont
		{Zhang}}]{Chen2009}%
	\BibitemOpen
	\bibfield  {author} {\bibinfo {author} {\bibfnamefont {W.-Q.}\ \bibnamefont
			{Chen}}, \bibinfo {author} {\bibfnamefont {K.-Y.}\ \bibnamefont {Yang}},
		\bibinfo {author} {\bibfnamefont {Y.}~\bibnamefont {Zhou}}, \ and\ \bibinfo
		{author} {\bibfnamefont {F.-C.}\ \bibnamefont {Zhang}},\ }\href {\doibase
		10.1103/PhysRevLett.102.047006} {\bibfield  {journal} {\bibinfo  {journal}
			{Phys. Rev. Lett.}\ }\textbf {\bibinfo {volume} {102}},\ \bibinfo {pages}
		{047006} (\bibinfo {year} {2009})}\BibitemShut {NoStop}%
	\bibitem [{\citenamefont {Goswami}\ \emph {et~al.}(2010)\citenamefont
		{Goswami}, \citenamefont {Nikolic},\ and\ \citenamefont {Si}}]{Goswami2010}%
	\BibitemOpen
	\bibfield  {author} {\bibinfo {author} {\bibfnamefont {P.}~\bibnamefont
			{Goswami}}, \bibinfo {author} {\bibfnamefont {P.}~\bibnamefont {Nikolic}}, \
		and\ \bibinfo {author} {\bibfnamefont {Q.}~\bibnamefont {Si}},\ }\href
	{http://stacks.iop.org/0295-5075/91/i=3/a=37006} {\bibfield  {journal}
		{\bibinfo  {journal} {EPL (Europhysics Letters)}\ }\textbf {\bibinfo {volume}
			{91}},\ \bibinfo {pages} {37006} (\bibinfo {year} {2010})}\BibitemShut
	{NoStop}%
	\bibitem [{\citenamefont {Yu}\ \emph {et~al.}(2013)\citenamefont {Yu},
		\citenamefont {Goswami}, \citenamefont {Si}, \citenamefont {Nikolic},\ and\
		\citenamefont {Zhu}}]{Yu2013-SC}%
	\BibitemOpen
	\bibfield  {author} {\bibinfo {author} {\bibfnamefont {R.}~\bibnamefont
			{Yu}}, \bibinfo {author} {\bibfnamefont {P.}~\bibnamefont {Goswami}},
		\bibinfo {author} {\bibfnamefont {Q.}~\bibnamefont {Si}}, \bibinfo {author}
		{\bibfnamefont {P.}~\bibnamefont {Nikolic}}, \ and\ \bibinfo {author}
		{\bibfnamefont {J.-X.}\ \bibnamefont {Zhu}},\ }\href
	{http://dx.doi.org/10.1038/ncomms3783} {\bibfield  {journal} {\bibinfo
			{journal} {Nat Commun}\ }\textbf {\bibinfo {volume} {4}},\ \bibinfo {pages} {2783} (\bibinfo {year}
		{2013})}\BibitemShut {NoStop}%
	\bibitem [{\citenamefont {Yu}\ \emph {et~al.}(2014)\citenamefont {Yu},
		\citenamefont {Zhu},\ and\ \citenamefont {Si}}]{Yu2014}%
	\BibitemOpen
	\bibfield  {author} {\bibinfo {author} {\bibfnamefont {R.}~\bibnamefont
			{Yu}}, \bibinfo {author} {\bibfnamefont {J.-X.}\ \bibnamefont {Zhu}}, \ and\
		\bibinfo {author} {\bibfnamefont {Q.}~\bibnamefont {Si}},\ }\href {\doibase
		10.1103/PhysRevB.89.024509} {\bibfield  {journal} {\bibinfo  {journal} {Phys.
				Rev. B}\ }\textbf {\bibinfo {volume} {89}},\ \bibinfo {pages} {024509}
		(\bibinfo {year} {2014})}\BibitemShut {NoStop}%
	\bibitem [{\citenamefont {Wen}(2015)}]{Wen2015}%
	\BibitemOpen
	\bibfield  {author} {\bibinfo {author} {\bibfnamefont {J.}~\bibnamefont
			{Wen}},\ }\href@noop {} {\bibfield  {journal} {\bibinfo  {journal} {Annals
				Phys.}\ }\textbf {\bibinfo {volume} {358}},\ \bibinfo {pages} {92} (\bibinfo
		{year} {2015})}\BibitemShut {NoStop}%
	\bibitem [{\citenamefont {Towns}\ \emph {et~al.}(2014)\citenamefont {Towns},
		\citenamefont {Cockerill}, \citenamefont {Dahan}, \citenamefont {Foster},
		\citenamefont {Gaither}, \citenamefont {Grimshaw}, \citenamefont {Hazlewood},
		\citenamefont {Lathrop}, \citenamefont {Lifka}, \citenamefont {Peterson},
		\citenamefont {Roskies}, \citenamefont {Scott},\ and\ \citenamefont
		{Wilkens-Diehr}}]{Towns2014}%
	\BibitemOpen
	\bibfield  {author} {\bibinfo {author} {\bibfnamefont {J.}~\bibnamefont
			{Towns}}, \bibinfo {author} {\bibfnamefont {T.}~\bibnamefont {Cockerill}},
		\bibinfo {author} {\bibfnamefont {M.}~\bibnamefont {Dahan}}, \bibinfo
		{author} {\bibfnamefont {I.}~\bibnamefont {Foster}}, \bibinfo {author}
		{\bibfnamefont {K.}~\bibnamefont {Gaither}}, \bibinfo {author} {\bibfnamefont
			{A.}~\bibnamefont {Grimshaw}}, \bibinfo {author} {\bibfnamefont
			{V.}~\bibnamefont {Hazlewood}}, \bibinfo {author} {\bibfnamefont
			{S.}~\bibnamefont {Lathrop}}, \bibinfo {author} {\bibfnamefont
			{D.}~\bibnamefont {Lifka}}, \bibinfo {author} {\bibfnamefont {G.~D.}\
			\bibnamefont {Peterson}}, \bibinfo {author} {\bibfnamefont {R.}~\bibnamefont
			{Roskies}}, \bibinfo {author} {\bibfnamefont {J.~R.}\ \bibnamefont {Scott}},
		\ and\ \bibinfo {author} {\bibfnamefont {N.}~\bibnamefont {Wilkens-Diehr}},\
	}\href {\doibase http://doi.ieeecomputersociety.org/10.1109/MCSE.2014.80}
	{\bibfield  {journal} {\bibinfo  {journal} {Computing in Science and
				Engineering}\ }\textbf {\bibinfo {volume} {16}},\ \bibinfo {pages} {62}
		(\bibinfo {year} {2014})}\BibitemShut {NoStop}%
\end{thebibliography}

%

\end{document}